\documentclass[sigconf]{acmart}

\acmConference[Decoding the Secrets of Machine Learning in Windows Malware Classification]{ACM Conference on Computer and Communications Security}{2023}{Copenhagen, Denmark}
\acmYear{2023}

\usepackage[english]{babel}
\usepackage{amsmath}

\usepackage{amssymb}
\usepackage{booktabs}
\usepackage{comment}

\usepackage{enumitem}
\usepackage{graphicx}
\usepackage{subcaption}
\usepackage{tabularx}
\usepackage{tikz}
\usepackage{xspace}
\usepackage{xurl}
\usepackage{array}
\usepackage{makecell}
\usepackage{multirow}
\setitemize{noitemsep,topsep=0pt,parsep=0pt,partopsep=0pt}

\renewcommand{\paragraph}{\vspace{1pt}\noindent\textbf}
\usepackage[colorinlistoftodos]{todonotes}

\newcommand{\revision}[1]{#1}

\newcommand{\summary}[2]{
\vspace{0.2cm}
\noindent \fcolorbox{gray}{white}{\begin{minipage}[c]{0.98\columnwidth}
  \aques{#1} #2
\end{minipage}}
\vspace{-.3cm}
}

\newcommand\cmBox[1]{
  \fbox{\lower0.38cm
    \vbox to 0.9cm{\vfil
      \hbox to 0.9cm{\hfil\parbox{0.45cm}{#1}\hfil}
      \vfil}%
  }%
}

\newcommand{\ignore}[1]{}
\usepackage{amsfonts}
\newcommand{\Y}{\checkmark}
\usepackage{pifont}
\newcommand{\N}{\hspace{1pt}\ding{55}}

\newcommand{\avclass}{AVClass2\xspace}
\newcommand{\trid}{trID\xspace}
\newcommand{\nummalware}{67,000\xspace}
\newcommand{\numbenign}{16,611\xspace}
\newcommand{\numfam}{670\xspace}
\newcommand{\vtag}[1]{\emph{#1}\xspace}

\newcounter{resnumber}

\newcommand{\rques}[1]{%
\textcolor{blue}{$\langle$\textbf{R#1}$\rangle$}%
}
\newcommand{\aques}[1]{%
\textcolor{blue}{$\langle$\textbf{A#1}$\rangle$}%
}

\begin{document}



\author{
  \begin{tabular}{c}
	Savino Dambra*\\
	Norton Research Group
  \end{tabular}
  \begin{tabular}{c}
	Yufei Han*\\
	INRIA
  \end{tabular}
  \begin{tabular}{c}
	Simone Aonzo*\\
	EURECOM
  \end{tabular}
  \begin{tabular}{c}
	Platon Kotzias*\\
	Norton Research Group
  \end{tabular}\\[-2pt]
  \begin{tabular}{c}
	Antonino Vitale\\
	EURECOM
  \end{tabular}
  \begin{tabular}{c}
	Juan Caballero\\
	IMDEA Software Institute
  \end{tabular}
  \begin{tabular}{c}
	Davide Balzarotti\\
	EURECOM
  \end{tabular}
  \begin{tabular}{c}
	Leyla Bilge\\
	Norton Research Group
  \end{tabular}
}

\date{}
\begin{abstract}
Many studies have proposed machine-learning (ML) models for malware detection and classification, reporting an almost-perfect performance. 
However, they assemble ground-truth in different ways, use diverse static- and dynamic-analysis techniques for feature extraction, and even differ on what they consider a malware family. 
As a consequence, our community still lacks an understanding of malware classification results: whether they are tied to the nature and distribution of the collected dataset, to what extent the number of families and samples in the training dataset influence performance, and how well static and dynamic features complement each other. 

This work sheds light on those open questions by investigating the 
\revision{impact of datasets, features, and classifiers} 
on ML-based malware detection and classification. 
For this, we collect the largest balanced malware dataset so far with 67k samples from 670 families (100 samples each), and train state-of-the-art models for malware detection and family classification using our dataset. 
Our results reveal that static features perform better than dynamic features, and that combining both only provides marginal improvement over static features. 
We discover no correlation between packing and classification accuracy, and that missing behaviors in dynamically-extracted features highly penalise their performance. 
We also demonstrate how a larger number of families to classify makes the classification harder, while a higher number of samples per family increases accuracy. 
Finally, we find that models trained on a uniform distribution of samples per family better generalize on unseen data. 

\end{abstract}

\title{Decoding the Secrets of\\ Machine Learning in Windows Malware Classification: \\A Deep Dive into Datasets, Features, and Model Performance}
\maketitle
\def\thefootnote{*}\footnotetext{These authors contributed equally to this work.}

\section{Introduction} \label{sec:introduction}

Modern Windows malware analysis has to cope with a large number of samples that
have been steadily increasing for two decades. 
In 2022, both the AV-TEST Institute and Kaspersky registered over 400,000 new malicious programs daily~\cite{kasperskyNewmalware,avtestNewmalware}. 
In order to counter such numbers, research and industry have begun to rely on Machine Learning (ML)-driven malware classification models. 
They can be applied over a large number of files and offer more flexible
classification mechanisms than signature-based methods.  Nevertheless, they have
to contend with human attackers' imagination, which consistently produces new
variants to fly under the radar.  At their core, ML techniques capture the
statistical correlation between training data and classification targets. As a
result, such statistics-based classification models lose their effectiveness
when going beyond the
knowledge encoded in the training data. Human attackers aware of this limitation
can thus always be one step ahead to choose attacks unseen in the training data,
in order to evade the detection of ML-based methods.  Moreover, ML-based
classification models are often performed in a pipeline~\cite{loi2021ITASEC,xiao2021csai,malinsight}. For example, given a
suspicious file, a typical ML pipeline should first figure out whether it is
malicious (\emph{binary classification}), and then find out whether it belongs to a
known family (\emph{family classification}). Even though these classification
tasks achieve high accuracy in previous literature~\cite{loi2021ITASEC,xiao2021csai,malinsight}, most of these works have
been carried out with unrealistic assumptions, mainly because of how the dataset
was constructed. 

In addition, a ground-truth of malware families is hard to obtain. Antivirus companies will not likely use the same name for the same
family. Although the CARO (Computer Antivirus Research
Organization) naming convention has been proposed to mitigate this issue, 
it still faces usage obstacles.
Scientific research tackled this problem and produced AVClass~\cite{avclass2}: 
given a list of AV labels (e.g., from a VirusTotal JSON report),
the tool returns the \emph{single most likely} family name.  However, even if AVClass
returns a single family name according to a consensus algorithm by default, it
can also output a ranking of all alternative family names. Thus, the problem is
that AVClass is often used to carry out studies using its default output as
ground truth, even though it is probabilistic in nature.

Moreover, while it is straightforward to collect a high number of samples for
popular families, collecting a large diverse malware dataset remains difficult
and time-consuming~\cite{anderson2018ember,smith2020mind,tesseract,lee2021android}.
In this work, we collect PE malware executables from the VirusTotal (VT) feed~\cite{vtfeedapi},
a real-time stream of JSON-encoded reports of samples submitted to VirusTotal.
Despite the appearance of more than 44M VT reports over a period of 
nearly three months and the collection of 227k samples from 13.8k families, 
only 780 malware families of those contain at least 100 samples.

To further complicate the matter, malware authors often use off-the-shelf
packers and protectors~\cite{miramirkhani2017spotless,maffia2021longitudinal}. 
Both modify a program to hinder its analysis while still preserving 
its original behavior. Based on their design, different malware that undergo the
packing or protection procedures may generate executables that share a highly
similar structure. This easily makes a ML classifier trained over these malware samples
overfit the packed or protected file structure, rather than capturing
its true malicious component.

Therefore, in this work, we put considerable effort to create four
heterogeneous datasets for a total of 118,111 samples to perform a large-scale
measurement study. Three of them are composed of malicious samples with
varying numbers of families, while the fourth contains benign samples.  We devoted
particular attention during the construction of the datasets, trying both to
reproduce the datasets usually used in research, but also considering
real-world scenarios typical of malware analysis. Such datasets allowed us to
create well-controlled experiments for studying how the effectiveness of
ML-based binary and family classification change under different testing
scenarios. 

Finally, there is also another crucial aspect that influences ML algorithms
that we further explored: \emph{feature} extraction. The methods by which one
can analyze executable files fall into two main categories, depending on what
facets one wants to study, namely \emph{static} properties and \emph{dynamic}
behavior; nonetheless, the previous two can also be \emph{combined}.  Since we
wanted to study existing ML state-of-the-art solutions and \textbf{not} design
new ones, we build our static and dynamic feature extraction approaches on what
was described in recent papers~\cite{AonzoUsenix2022,aghakhani2020malware}.
Therefore, this means that we have statically analyzed and dynamically executed
in a sandbox more than a hundred thousand samples were used in this study.

\vspace{0.3em}

Our work contributes by answering the following research questions
for both binary and family classification tasks:

\noindent \rques{1} \textbf{How do static, dynamic, and combined models perform
on different malware families/classes in binary and family classification?}

\noindent\rques{2} \textbf{On which families and classes of malware does each model fail to produce accurate classification? }

\noindent\rques{3} \textbf{What is the contribution of static and dynamic feature
classes to the classification performance and does their contribution change when
joining the two sets?}

\noindent\rques{4} \textbf{Does the presence of off-the-shelf packers and protectors bring harm to classification accuracy? }

\noindent \rques{5} \textbf{Do missing feature values in the runtime behaviors
negatively impact the classification results?}

\noindent\rques{6} \textbf{Is the AVClass2 confidence score correlated with
\mbox{ML-based} decisions?}

\noindent\rques{7} \textbf{How does the training dataset construction strategy
affect the model performance?}  

\noindent\rques{8} \textbf{How does the ML-driven malware classifier perform
over the families unseen in the training data?}

\section{Dataset Collection}
\label{sec:datasets}

To conduct our experiments we collected 118,111 Windows PE32
executables, divided in four datasets, 
as summarized in Table~\ref{tbl:datasets}.
This section describes the process for building those datasets.


\begin{table}[t]
\small
\centering
\caption{Dataset summary}
\begin{tabular}{l|r|rrr}
\hline
\textbf{Dataset}  & \textbf{Samples} & \textbf{Families} \\
\hline
Malware Balanced ($M_B$)  & \nummalware & \numfam \\
Benign           ($B$) & \numbenign & - \\
Malware Unbalanced ($M_U$)   &  18,000 & 1,500 \\
Malware Generic ($M_G$)  &  16,500 & - \\
\hline
All &  118,111 & - \\
\hline
\end{tabular}
\label{tbl:datasets}
\end{table}

\subsection{Malware Samples}
\label{sec:malcollect}

We collect PE malware executables from the VirusTotal (VT) feed~\cite{vtfeedapi}.
The VT feed is a real-time stream of JSON-encoded reports. 
Each report contains the analysis results of a sample submitted to VirusTotal --
including file hashes, filetype, size, and 
the detection labels assigned by a large number of antivirus (AV) engines.
These reports are generated both by new samples submitted by VT users, as well as by
user-requested re-analysis of files already known to VT.  Samples in the feed
can be of various file types (e.g., PE, APK, PDF),
but our collection focuses on Windows PE executables. 
Samples that appear in the feed can be downloaded
within 7 days from the moment they appear in the feed.

We want our dataset to be as diverse as possible in terms of the number of families,
but also to be balanced,
so that no malware family is over-represented or under-represented.
Our initial target was to collect 1,000 malware families
with a hundred samples each.
The threshold of 100 samples per family was chosen to have enough samples per family to performing
multi-class classification experiments,
taking into account that samples are split into
60\% training, 20\% validation, and 20\% testing.
However, due to the collection, filtering, and reclassification process
described below, we ended up with
\numfam families satisfying that threshold,
as shown in Table~\ref{tbl:datasets}.

To the best of our knowledge, 
this is the most diverse labeled malware dataset in terms of families
to date.
The most recent dataset was Motif~\cite{motif} with 454 families. 
While the number of families in Motif is also large, 
it is 21 times smaller than our balanced dataset with 
3,095 samples, and is unbalanced with a median of three samples per family. 
Only one family in Motif reaches 100 samples and 
29\% of the families have only one sample. 
Such a small number of samples for most families does not allow 
building an accurate multi-class classifier, 
as we will show in our evaluation.

\paragraph{Initial collection from VT feed.}
We collected reports and samples from the VT feed
for 82 non-consecutive days between August 2021 and March 2022.
We only retained reports of samples 
detected by at least one AV engine, 
and with a \trid~\cite{trid}
filetype identification field (available in the report) equal to 
`32-bit non-installer PE executable'.
We excluded 64-bit PE executables,
dynamic-link libraries (DLLs), and
executables generated by popular installer software (e.g., NSIS, InnoSetup).
These restrictions are placed by our dynamic analysis sandbox,
described in Section~\ref{sec:sandbox},
which currently does not support running
64-bit PE executables or DLLs, and does not interact with GUIs in order to
complete the installation of other programs.
%
However, an analysis of the whole VT feed during the 82 collection days shows that
from all malicious PE samples in the feed,
87.6\% are 32-bit executables,
8.2\% are DLLs (32-bit or 64-bit),
3.9\% are 64-bit executables,
and the remaining 0.3\% are other PE types (e.g., OCX, CPL, SCR).
%

The retained reports are fed to the \avclass malware labeling
tool~\cite{avclass2}, which outputs the most likely family name 
for the sample as well as a confidence factor that captures the number of 
AV engines assign that family to the sample 
(after removing duplicates due to AV engines that copy each other).
For each family reported by \avclass, our system downloaded
100 distinct samples.
Each downloaded sample was then checked again to exclude any remaining non-32-bit PE
executables and installers that were missed by \trid.  In
particular, samples are removed if their PE header does not indicate 
they are 32-bit executables, or if they are detected as installers
using public Yara rules by Avast~\cite{retdecYara}.
As stated, our initial target was to collect 1,000 malware families
with 100 samples each.
However, when this target was reached, many other families had been
collected with less than 100 samples,
resulting in an initial dataset of 239,417 PE32 malware samples from 23,555 families.

\paragraph{Reclassification and family filtering.}
The AV labels of a sample may change over time
as AV vendors refine their detection rules.
These label changes may in turn change the family that \avclass
outputs for a sample.
To account for such changes, we re-collect the updated VT report for our
samples 54 days after the end of our collection process,
and feed the new reports to \avclass to obtain the (possibly) updated family.
From the 239,417 samples, 9.7\% (23,171) were at this point re-classified as a
different family.
\avclass uses a taxonomy to identify a wide range of non-family tokens that may appear in the AV
labels. These include 
file properties (e.g., \vtag{FILE:packed:asprotect}, \vtag{FILE:exploit: gingerbreak}), 
malware classes (e.g., \vtag{CLASS:virus}, \vtag{CLASS:worm}),
behaviors (e.g., \vtag{BEH:ddos}, \vtag{BEH:filedelete}), 
and generic tokens (e.g., \vtag{GEN:malicious}, \vtag{GEN:behaveslike}).  However, the
\avclass taxonomy is assumed to be incomplete by design~\cite{avclass2}.  Thus,
it may output a label for a sample that does not correspond to a real family,
but rather to a previously unknown instance of the above categories.
To address this issue, we manually inspected the collected family labels and
conservatively filtered out any labels that may not correspond to real family names.
This step identified 86 likely non-family tokens
not in the \avclass taxonomy, such as
\vtag{gametool}, \vtag{testsample}, \vtag{nsismod}, \vtag{dllinject}, and
\vtag{processhijack}.  We also removed random-looking labels (e.g.,
\vtag{005376ae}) that \avclass failed to filter.
As a byproduct of our effort, we will contribute our extended \avclass taxonomy 
to the open-source \avclass project.

After reclassification and family filtering, the dataset contained
227,296 samples from 13,894 families, out of which 
780 families had at least 100 samples.
Thus, despite examining more than 44M VT reports over a period of 
nearly 3 months, we were unable to reach our goal of 
1,000 families with 100 samples.
This illustrates the difficulty of building a diverse malware dataset.




\paragraph{Feature filtering.}
We performed static and dynamic feature extraction (as detailed in
Section~\ref{sec:methodology}) for all samples of the 780 families with at least 100 samples.
This required executing each sample in a sandbox to obtain a behavioral report.
\revision{We discarded 122 samples for which the
static feature extraction pipeline failed. The failure reasons 
were corrupted headers (26 binaries), 
empty output from the disassembler 
probably due to obfuscation techniques (95 samples), and the absence of the
entry point in one binary. 
}
We also discarded samples that did not exhibit any runtime behavior,
and sub-sampled families to keep only
100 samples each. The result is a balanced dataset (hereinafter $M_B$)
that contains \nummalware samples from \numfam families.
According to \avclass,
those families belong to 13 malware classes:
36\% (282) of the families are classified as grayware
(including its adware subclass),
15\% (120) as downloaders,
11\% (87) as worms,
10\% (78) as backdoors,
5\% (41) as viruses,
and the remaining 23\% (62) includes ransomware, rogueware,
spyware, miners, hacking tools, clickers, and dialers.





\paragraph{Dataset statistics.}
Over 93\% of the samples in the $M_B$ dataset 
are detected by at least 20 AV engines, 
while only 0.3\% 
have a VT score less or equal to 3. 
It is worth noting that the minimum number of detections for samples in
the dataset is two since \avclass requires at least two AV engines to
assign a label to a sample.

Samples on the VT feed can be new (i.e., collected and scanned for the very
first time by VT) or resubmitted (i.e., first submitted in the past but
re-scanned on the day they were collected).
We compute the freshness of samples in the $M_B$ dataset
as the number of days between a sample's collection date and
its VT first seen date.
We observe that 
53.4\% of the samples were collected within a day of being first observed by VT,
7.6\% within a year,
and 37.8\% are old samples first seen over one year before our study.

\paragraph{Packer and protector detection.}
To hamper analysis, malware authors may use packers that compress 
a sample and de-compress it at runtime, as well as more sophisticated 
protectors that may combine different obfuscations such as 
packing, encryption, and code virtualization.
To evaluate the impact of packers and other protectors on 
malware classification, 
we determine whether a sample uses an off-the-shelf 
packer or protector by using 
the signature-based Detect It Easy (DIE)~\cite{detectiteasy} tool,
as well as the well-maintained Yara rules of Avast RetDec~\cite{retdecYara}.
Overall, 22\% of the samples in $M_B$ use a packer or protector. 
The most popular packer is  
\texttt{upx} detected on 14.0\% the samples, followed by
\texttt{aspack} (3.2\%) and
\texttt{pecompact} (1.0\%).
The most popular protectors are
\texttt{vmprotect} (1.9\%) and 
\texttt{asprotect} (0.4\%).

\subsection{Testing Datasets}
\label{sec:maltest}

We create two other disjoint malware datasets,
which we use in Section~\ref{sec:experimental} to 
test the ability of ML classifiers to generalize beyond the $M_B$ dataset 
they were built upon. 
The first dataset, referred as Malware Unbalanced (or $M_U$) in Table~\ref{tbl:datasets}, contains 18K samples from 1.5K families.
These samples were part of the initial VT feed collection, 
passed the filtering and re-classification steps, 
but their families never reached the threshold of 100 samples and 
thus were excluded from $M_B$.
All samples are detected by at least 20 AV engines and none of the samples 
nor their families are part of $M_B$.

The second dataset, Malware Generic ($M_G$), contains 16.5K samples 
for which AVClass2 was unable to output a family,
due to AV engines using only generic labels. These
samples were separately collected from the VT feed 
between June 23rd and July 6th 2022 and underwent the filtering steps 
to keep only 32-bit non-installer PE executables.
All samples are detected by at least 20 AV engines and 
none of the samples are part of $M_B$.

\subsection{Benign Samples}
\label{sec:benigncollect}

Building a benign dataset by just relying on the number of AV detections 
in the VT report is prone to errors
due to the presence of malicious files that are still unknown to AV engines.
Therefore, we took a more conservative strategy and decided to build a benign dataset
by using a fresh installation of all the community-maintained packages
of Chocolatey~\cite{chocolatey} 
(which undergo a rigorous moderation review process to avoid pollution)
in a clean machine running Windows 10.
After each package was installed, we extracted all the executable files
present on the hard disk, which may correspond to Windows system files or
third-party publishers.

We exclude files that are not 32-bit PE executables and those
with more than three detections on VT.
This allowed us to discard borderline cases,
i.e., benign files with characteristics very similar to malware, like
hacking and scanning tools.
Using this procedure we collected a dataset $B$ of \numbenign benign samples.
The code signatures of those samples indicate a large diversity of publishers with
over 1.4K different signers -- including both small companies and 
large software publishers such as Microsoft, Oracle, and Google.

\section{Methodology} \label{sec:methodology}

Our work aims to answer the 8 research questions raised in 
the introduction. 
Notably, we aim to explore the performances of ML-driven malware classifiers
that use features extracted statically, dynamically, or a combination of both
with varied coverage of malware families and changed volumes of training
samples.  Developing novel ML-based malware
classification models is beyond the scope of our study. Instead, we focus on
discussing and evaluating the analysed issues using state-of-the-art ML models
for malware classification. 
\revision{As explained next, we use features presented in previous
works~\cite{aghakhani2020malware,mtnet,malinsight,AonzoUsenix2022}. 
This imposes a limitation as other features could provide better results.
}
%

\begin{table}[t]
\centering
\small
\caption{Feature classes used in the classifiers.}
\begin{tabular}{l|l|l|r}
\hline
 \textbf{ID}
 &\textbf{Class}
 &\textbf{Extraction}
 &\textbf{Features} \\ 
\hline
s-headers & PE headers & static & 29 \\ 
s-sections & PE sections & static & 590 \\ 
s-file & File Generic & static & 2 \\ 
s-dll & DLL imports & static & 131 \\ 
s-imports & API imports & static & 3,732 \\ 
s-strings & Strings & static & 10,402 \\ 
s-bytegrams & Byte n-grams & static & 13,000 \\ 
s-opcodegrams & Opcodode n-grams & static & 2,500 \\ 
\hline
d-network & Network activity & dynamic & 438\\
d-file & File activity & dynamic & 60,555\\
d-mutex & Mutexes used & dynamic & 7\\
d-registry & Registry operations & dynamic & 60\\
d-service & Services activity & dynamic & 736\\
d-process & Process activity & dynamic & 28,198 \\
d-thread & Thread actitivy & dynamic & 7 \\
\hline
\end{tabular}
\label{tbl:features}
\end{table}

\subsection{Static Features} \label{sec:features-static}

Hojjat et al.~\cite{aghakhani2020malware} performed a literature review to
identify the static features that carry the most useful information for binary
classification. We implement their feature extraction methodology to extract the
same classes of static features.
\revision{Similar to Hojjat et al.~\cite{aghakhani2020malware}, 
we do not attempt to unpack the executables and
perform the same feature extraction regardless of whether the files are packed or not.}

The upper half of Table~\ref{tbl:features} summarizes the static feature 
classes (prefixed by \texttt{s-}).
The \emph{s-headers} class captures 29 integer features 
(Section \ref{sec:peHeaders-peSections} in the Appendix)
from the \textit{Optional} and \textit{COFF} headers of the
executable~\cite{microsoftPE}.
The \emph{s-sections} class captures 590 Boolean features 
from each section in the executable (Section \ref{sec:peHeaders-peSections} in the Appendix).
\revision{
The \emph{s-file} features capture the file size in bytes and the whole
file Shannon entropy~\cite{lyda2007using}.}

\revision{
For the remaining 5 feature classes
the exact number of features may differ from 
those reported by~\cite{aghakhani2020malware} because they undergo 
a dataset-dependent feature selection step that retains only the features that
show variability or that provide higher information gain
(IG)~\cite{quinlan1986induction}.
For instance, in \emph{s-bytegrams} and \emph{s-opcodegrams},
the selection process enumerates all values observed in the 
validation set (20\% of samples in $M_B$), 
excludes rare values appearing in less than 1\% of the samples, 
computes IG, 
uses the elbow method to identify a threshold value for IG, and
only retains features with at least that threshold IG.
As in~\cite{aghakhani2020malware}, for \emph{s-dll}, \emph{s-imports}, and \emph{s-strings}, 
the selection process only excludes rare values, 
but does not select an IG threshold.
} 

\revision{
The \emph{s-dll} and \emph{s-imports} class contain  
Boolean features extracted from the import table (imported libraries
in case of \emph{s-dll} and imported functions for \emph{s-imports}).
We extracted 637 unique libraries and 28,667 
functions and retained only those that appear in at least 1\% of the files 
in the validation set, reducing the number to 131 DLLs and 3,732 library functions.
Similarly, for the \emph{s-strings} class, we extracted 106,352,885
strings of at least 4 characters, filter those that appear in over 1\% of the
files, and kept 10,402 Boolean features capturing whether the string appears or
not in the binary.
The \emph{s-bytegrams} class captures the presence of selected 
4-grams, 5-grams, and 6-grams. 
As proposed in~\cite{aghakhani2020malware}, to keep memory usage manageable, 
the selection process for this feature class is performed on 1,000 randomly 
chosen files from $M_B$, instead of the full validation dataset.
From the 1,363,150,788 \emph{s-bytegrams} extracted, 
the selection retained the 13,000 features with the highest IG
(Section~\ref{sec:IG} in the Appendix).}
\revision{The \emph{s-opcodegrams} class captures 1-gram, 2-grams and 3-grams
from the sequence of opcodes disassembled using Capstone~\cite{capstone}.
Given an initial set of 255,812 opcode n-grams, we computed the TF-IDF
and used the elbow method on the IG distribution 
to retain the top 2,500 float features (Section~\ref{sec:IG} in the Appendix).}

\subsection{Sandbox} \label{sec:sandbox}

We have built a sandbox for executing malware using the best practices proposed
by previous
works~\cite{rossow2012prudent,miramirkhani2017spotless,maffia2021longitudinal,yong2021inside}.
We configured a Windows 10 Pro 32-bit virtual machine (VM) with 2 CPUs (Intel
Xeon Platinum 8160 @ 2.10GHz) and 2 GiB of RAM.  We installed popular apps and
populated the file system with common file types to resemble a legitimate
desktop workstation as suggested by Miramirkhani et
al.~\cite{miramirkhani2017spotless}.  Malware runs on clones of this VM
orchestrated using Proxmox VE~\cite{proxmoxve}.  To improve performance, we
stored all virtual disk images and VM snapshots in a RAM disk.
As recommended by Rossow et al.~\cite{rossow2012prudent}, each machine runs on
its isolated local network with full Internet access through an ADSL line of
our institution dedicated to this purpose.
Recent works have measured that 40\%--80\% of modern malware use at least one
evasive technique~\cite{maffia2021longitudinal,galloro2022systematical}.  To
limit the impact of such evasions, we base our analysis on the Intel PIN-based
\emph{JuanLesPIN} tool~\cite{JLP,maffia2021longitudinal}, which handles common
evasive techniques, thereby increasing the likelihood that malware detonates.
Unfortunately, it does not support 64-bit Windows executables, so we focus on
32-bit malware. We modified \emph{JuanLesPIN} to monitor Windows APIs
responsible for network, processes, services, registry, mutexes, file system,
and DLL loading. Finally, 
we tested our analysis environment with the Al-Khaser~\cite{alkhaser} tool to
confirm that our sandbox could not be identified.
To measure the overhead introduced by our analysis system we executed 1,000 malware samples
randomly chosen among those that: (i) terminate the execution, (ii) use at least
one evasive technique, and (iii) \textbf{detonates} according to the threshold
proposed in~\cite{kuechler2021}, 
i.e., the sample calls at least 50 Windows APIs.
We measured their execution time with and without instrumentation by observing a
percentage increase of $\mu=125$, $\sigma=31$, $min=26$, $med=106$, $max=206$.
This overhead is in line with that in~\cite{maffia2021longitudinal}.
Kuechler et al.~\cite{kuechler2021} recently showed that the amount of code
executed by malware samples plateaus after two minutes, and little additional
information can be obtained thereafter. 
Thus considering the overhead mentioned above, we took a conservative approach and ran
each sample for up to five minutes.

\subsection{Dynamic Features} \label{sec:features-dynamic}

We extract 7 classes of dynamic features from the API calls (including their
arguments) invoked by the malware during execution in the sandbox.  The features
were chosen to cover those used in previous works that built classifiers from
malware executions (e.g.,~\cite{mtnet,malinsight,AonzoUsenix2022}).

The lower half of Table~\ref{tbl:features} summarizes the 7 dynamic feature
classes (prefixed by \texttt{d-}).
%
Categorical features such as filenames and domains are one-hot encoded to
Boolean features.  To encode each feature, we count all its possible values and
exclude those appearing less than five times in the training set. 
%
The \emph{d-network} class (438 features) captures the HTTP, TCP, and UDP
traffic.  Of those, 430 features capture unique domains contacted by the malware
and HTTP User-Agent strings used; three count the number of HTTP requests, TCP
connections, and UDP pseudo-sessions; and 5 randomness-related 
features capture the mean/median/min/max/std likelihood of domain names and URLs
contacted according to a recently proposed Markov Chain
model~\cite{AonzoUsenix2022}.
%
The \emph{d-file} class features (60,555) capture the name and extension of
60,547 files created or accessed by the malware, the number of files read,
written, and deleted; and 5 capture the randomness of the filenames. 
%
The \emph{d-mutex} class features (7) capture the number of mutex objects
created and the randomness of the mutex names.
The \emph{d-registry} class features (60) capture 55 unique registry keys
written, and the count of registry keys created, opened, read, written, and
deleted.
The \emph{d-service} class features (736) capture the count, randomness, and
names of services and service managers created, started, and halted. 
The \emph{d-process} class features (28,198) capture the count of processes
created, processes terminated, and shell commands invoked, as well as 28,195
unique process names.
The \emph{d-thread} class features (7) capture the number of the threads opened,
created, resumed, terminated, and suspended, as well as the number of the
interactions with the context of a given thread and the number of asynchronous
procedure calls (APC) queued to a thread.  The last two features help capture
suspicious behaviors.

\paragraph{Missing features.}
When a dynamic feature cannot be computed (e.g.,due to lack of activity), 
we assign them default place-holder values that do
not belong to the domain of the features.
We refer to such features as \emph{missing} features.  For example, if a sample
has no file system activity, we cannot compute the \emph{d-file} filename
randomness features. As a result, the 5 statistical features related to the
randomness of the file names are thus not available.
\revision{We perform dynamic feature extraction only over detonated 
malware samples 
(i.e., those that called at least 50 APIs as defined in \ref{sec:sandbox}), 
but even for detonated samples, there are still missing observations of feature values.}
To facilitate the analysis of the impact of the missing features, we
define the \emph{feature missing rate} (FMR) of a malware family as the fraction
of family samples that have missing values in the file, registry, service, and
process features
(which, among the seven dynamic features classes we consider, are the most
relevant for classification according to
Table~\ref{tbl:binary_and_multiclass_FeatImportance}).  Missing values over
\text{all} these four feature classes considerably degrades both the amount and
quality of useful information available to the classifier.  According to our
analysis, over 54\% of the malware families studied in our work contain on
average 77\% of the malware samples per family with missing feature values in
these four dynamic feature classes. Missing observations can negatively impact
ML classifiers by overfitting the data and reducing the model's accuracy.
Recently, Aonzo et al.~\cite{AonzoUsenix2022} showed that classifier models tend
to focus on static features, rather than dynamic ones, precisely because static
features are rarely missing.
%
In Section~\ref{sec:best_and_worst} we analyze the impact of missing features in
the classification results.

\subsection{Models} \label{sec:models}

We train multiple models to capture different axis: 
classification task (i.e., binary or family classification), 
features (i.e., static, dynamic, combined), 
\revision{classifiers (i.e., Random Forest, XGBoost),} 
dataset construction (i.e., distribution of families in training
dataset), and a different number of families and samples.

\paragraph{Classification task.} We build models for binary and family
classification tasks. The binary classification models detect whether a
given sample is malicious (positive class) or benign (negative class). The
family classification models identify the family of a given malicious
sample, that is, there is one class per malware family and no goodware
class.  We prefix the name of a model with \emph{binary-} or \emph{family-}
to indicate the classification task.

\paragraph{Features.} We build models that use all static features, all
dynamic features, and all combined features (i.e., all static and all
dynamic).  The name of a model includes \emph{-static-}, \emph{-dynamic-},
or \emph{-combined-} to indicate the features used. 

\paragraph{Classifiers.}
\revision{
Given a large number of ML classifiers, it is not possible for us to 
systematically evaluate all of them. 
In our experiments we selected \textit{Random Forest} and \textit{XGBoost} because
they are consistently among the best-performing classifiers
evaluated in previous works
(summarized in Table~\ref{tbl:related} and Section~\ref{sec:related}).
Moreover, being tree-based, they are easier to interpret, they allow 
direct analysis of feature importance, and they are also intrinsically 
capable of handling both categorical features
(e.g., unique filenames accessed during execution) and continuous features
(e.g., filename mean randomness).
We also considered neural networks, but discarded them because
to achieve good performance they require larger training datasets 
(e.g., $\ge 400k$ samples in~\cite{malconv}).
It was not clear whether we could build a balanced family dataset of 
the required size. 
In addition, there exist many potential neural architectures to evaluate and
their training times are longer,
which is critical given the large number of models we evaluate.
%
}

\paragraph{Dataset construction.} For the binary classification task, we
experiment with two ways of building our dataset, namely \emph{uniform} and
not \emph{nonUniform}.
The \emph{uniform} approach builds datasets that balance the number of
goodware and malware, using a sampling-with-replacement approach, as
follows. We uniformly select from each family in $M_B$ a number of samples
so that the total number of malicious samples matches the size of the benign
dataset (i.e., each family in $M_B$ provides 24--25 samples for a total of
16,611 malware samples).  We repeat the process five times avoiding
repetitions (i.e., each time selecting a different set of malware samples
from each family in $M_B$), to completely cover all the malicious samples
in each family.  These steps produce 5 balanced datasets. 
Each dataset is split into 60\% of samples for training, 20\% for
validation (i.e., selecting the classifier hyper-parameters), 
and 20\% for testing.
To evaluate a model, for each of the five datasets, we perform a 10-fold
cross validation to ensure that all the samples equally contribute to the
training and testing datasets. We report average results across the five
rounds and their respective folds. Thus, obtaining the accuracy results from
one model requires us to train and test 50 times.

The \emph{nonUniform} approach replicates the unbalanced distribution of
samples per family in the Motif dataset~\cite{motif}.  The motivation for
this dataset is to study whether the family distribution in the training set
of a binary classification task (where family labels are not used) affects
the detection accuracy.
In Motif, 29\% of families have only one sample, 41\% have 2-5 samples, 12\%
6-10, 10\% 11-20, 4\% 21-30, 2\% 31-40, 1\% 41-50, and 1\% has over 142
samples.  We replicate this distribution on the 670 families in $M_B$.  For
example, we select one sample from 29\% (randomly-chosen without
replacement) of the 670 $M_B$ families and 142 samples from one
randomly-chosen family. 
The resulting dataset comprises all 16,611 benign samples and 4,821 samples
from 670 families that follow the per-family sample distribution in Motif.

\paragraph{Number of families and samples.} To measure the impact that the
number of families to classify and the available samples for each family
have on the results, we build multiple ML-based classifiers for the family
classification task by uniformly sampling 70, 170, 270, 370, 470 and 570
families from the total 670 families. For each of them, we also experiment
with a version trained and tested on 50, 60, 70, and 80 malware samples for
each family. As indicated above, we have 20\% samples used as the validation
data.  Therefore, at maximum, there are 80 malware samples for training and
testing use.

\section{Experimental study} \label{sec:experimental}

This section presents the results of the experiments we conducted to answer the
research questions presented in the introduction.
We have adopted the following structure for ease of reading: 
the reader will find the discussion to \rques{x} in Section~\ref{sec:experimental}.x 
and a summary with the answer \aques{x} at the end of each subsection.


\begin{table}[t]
\footnotesize
\centering
\caption{\revision{Overall classification results using Random Forest.}}
\label{tbl:overallResults}
\begin{tabular}{llcccc}
\toprule
	\multirow{2}{*}{\textbf{Task}} &
	\multirow{2}{*}{\textbf{Features}} & \multirow{2}{*}{\textbf{Precision}} &
	\multirow{2}{*}{\textbf{Recall}} & \multirow{2}{*}{\textbf{F1-score}} &  \textbf{Families with} \\
	&  & & &  & \textbf{100\% accuracy} \\
\midrule
	Binary		&		Static		&	0.956	&	0.957	&	0.957		&		242 (36.12\%)\\	
	Binary		&		Dynamic		&	0.945		&	0.892		&	0.926		&		465 (69.40\%)\\	
	Binary		&		Combined	&	0.963	&	0.934	&	0.948		&		450 (67.16\%)\\	
	\midrule                         	 	                     
	Family		&		Static		&	0.856 	&	0.850	&		0.848		&		68 (10.15\%)\\	
	Family		&		Dynamic		& 0.734	& 0.708	&		0.704	&		114 (17.17\%)\\	
	Family		&		Combined	&	0.874	&	0.867 	&		0.865		&		138 (20.60\%)\\	

\bottomrule
\end{tabular}
\end{table}

\subsection{Overall Classification Results} 
\label{sec:classification-results}

In this section, we examine
how static, dynamic, and combined features impact
binary and family classification.
\revision{We first discuss the results using Random Forest and then 
discuss the XGBoost results.
Table~\ref{tbl:overallResults} summarizes the accuracy results 
using Random Forest.} 
The results correspond to the uniform dataset construction approach. 
Each line in the table reports the averaged precision, recall, and F1 score of 
10-fold cross validation. 
It also reports the fraction of malware families with 100\%
family-wise accuracy.
In binary classification, 100\% family-wise accuracy for a family
denotes that the family can be perfectly differentiated from goodware.
In family classification, 100\% family-wise accuracy instead means that 
samples from a malware family are not misclassified as another malware family.
%

The static features achieve a higher F1 score than the dynamic features 
in both binary and family classification.
However, the fraction of perfectly classified malware families is higher 
for dynamic features, suggesting that  
dynamic features work very well for some malware families, but poorly on others.
The combination of static and dynamic features brings 
marginal improvements in F1 score over static-only features. 
It improves it by 1\% for family classification, 
but decreases it by 2\% for binary classification.
On the other hand, adding dynamic features increases the 
percentage of perfectly classified families over the static case, 
although for binary classification the fraction reduces compared to 
dynamic-only features.
The accuracy reduction with more features might seem counter-intuitive,
but it can happen when the two feature sets are not independent and
bring different strengths and weaknesses that lead
to mistakes on different parts of the input space. It is well known as the curse-of-dimensionality in machine learning \cite{Trunk1979pami}. 
Adding more features does not necessarily improve the overall accuracy,  more features may bring unexpected variance and noise into the classification module \cite{Lip2012}. 
%

Our results may raise concerns about the value of dynamic analysis. 
On the one hand, dynamic features outperform static features for a 
fraction of families, significantly raising the number of perfectly classified 
families (e.g., nearly doubling it for binary classification).
This confirms the value of dynamic analysis, for example when researchers are interested 
to build behavioral signatures for specific malware families. 
On the other hand, the overall impact of adding dynamic features 
to static features is unclear. 
This might be the consequence of malware families for which dynamic features
do not work well, because of intrinsic properties of the 
malware family (or malware class), 
but also because the sandbox might fail to stimulate samples adequately 
(e.g., due to evasion techniques or to the lack of a live 
command-and-control server). 
%
\revision{Adding dynamic features to the models may still provide other 
benefits. 
For example, recent work has shown that dynamic features are preferred 
by humans for interpretability~\cite{AonzoUsenix2022}. 
Furthermore, dynamic features can increase the robustness of the model, 
making it more resilient to obfuscations designed to hamper static analysis.}

\begin{table}[t]
\footnotesize
\centering
	\caption{\revision{Overall classification results for XGBoost.}}
\label{tbl:overallResultsXGboost}
\begin{tabular}{llccc}
\toprule
	\textbf{Task} & \textbf{Features} & \textbf{Precision} & \textbf{Recall} & \textbf{F1-score} \\
\midrule
	Binary		&		Static		&	0.907	&	0.902	&	0.904 \\	
	Binary		&		Dynamic		&	0.978	&	0.820	&	0.892 \\	
	\midrule                         	 	                     
	Family		&		Static		&	0.705	&	0.690	&   0.697 \\
	Family		&		Dynamic		&	0.720	&	0.689	&	0.691 \\	
\bottomrule
\end{tabular}
\end{table}

\paragraph{XGBoost.}
\revision{Table~\ref{tbl:overallResultsXGboost} shows the classification 
results using XGBoost for static and dynamic features.
The results correspond to the uniform dataset construction approach 
and 10-fold cross validation.
Similar to Random Forest, 
the static features achieve higher F1 score than the dynamic features 
in both binary and family classification, 
although the advantage of static over dynamic is smaller in this case.
Compared to Random Forest classifiers, 
XGBoost classifiers have lower F1-score for 
both binary classification (4.4\% lower) and
family classification (8.2\% lower).
We failed to run XGBoost on the combined features 
due to XGBoost's higher memory consumption, 
which becomes a bottleneck given the large number of features (roughly 100k)
in the combined model. Although the marginal contribution of a combined feature
set was shown by using a single architecture, we point out that the combined analysis 
is in nature a feature-level information fusion whose utility is not dependent on the
classifier architecture but on how complementary the two feature sources are
between each other.
Since Random Forest classifiers have higher accuracy, 
and they also run faster than XGBoost classifiers
while consuming less memory, 
we use Random Forest classifiers in the rest of our evaluation.
}

\paragraph{Time-aware experiments.}
\revision{To avoid the temporal bias that cross-validation may introduce,
Pendlebury et al.~\cite{tesseract} suggested to split training samples
into temporal bins.
However, since our dataset only contains 100 samples per family, 
the individual bins would be too small and thus we decided to not perform temporal binning.
Instead, in Section~\ref{sec:singleton_and_unseen} 
we perform a separate out-of-distribution (OOD) evaluation 
with unseen families and singletons not present in the training dataset,
which addresses the main bias that cross-validation introduces.}

\summary{1}{
For both binary and family classification tasks,
models trained on static features alone provide
higher accuracy than the models trained only on dynamic features.
The latter is able to perfectly classify more families, 
but perform poorly on others, producing an overall lower 
classification accuracy.

Adding dynamic features on top of the 
static features brings marginal \revision{accuracy} improvement 
for family classification 
and even negatively affects binary classification.
\revision{On the other hand, dynamic features may offer benefits for 
model robustness and interpretability.}
}

\subsection{Hard-to-Detect Malware}
\label{sec:best_and_worst} 

\begin{table}[t]
\setlength\tabcolsep{3pt}
\footnotesize
\centering
	\caption{Classification accuracy for malware classes.}
\label{tbl:overallResultsGrouped}
\begin{tabular}{l|ccc|ccc}
\toprule
	\multirow{2}{*}{\textbf{Class}} &  \multicolumn{3}{c|}{\textbf{Binary
	class. Recall}} & \multicolumn{3}{c}{\textbf{Family class. F1 score}} \\
	&  \textbf{Static} & \textbf{Dyn.} & \textbf{Comb.} & \textbf{Static}
	& \textbf{Dyn.} & \textbf{Com.} \\
\midrule
Adware     &	0.905	& 0.915	& 0.981 &   0.926 &	0.761	&	0.925	\\
Backdoor   &	0.966	& 0.943	& 0.996 &   0.830 &	0.730	&	0.838	\\
Clicker    &	0.971	& 0.929 & 1.000 &   0.817 &	0.692	&	0.821	\\
Dialer     &	0.994	& 0.875	& 1.000 &   0.988 &	0.888	&	0.984	\\
Downloader &	0.974	& 0.899	& 0.996 &   0.864 &	0.695	&	0.874	\\
Grayware   &	0.932	& 0.895	& 0.986 &   0.832 &	0.675	&	0.852	\\
Miner      &	0.989	& 0.972	& 0.999 &   0.927 &	0.807	&	0.962	\\
Ransomware &	0.967	& 0.945	& 0.997 &   0.839 &	0.580	&	0.853	\\
Rogueware  &	0.984	& 1.000	& 0.992 &   0.616 &	0.401	&	0.663	\\
Spyware    &	0.972	& 0.829	& 0.998 &   0.869 &	0.704	&	0.879	\\
Tool       &	0.992	& 0.929	& 1.000 &   0.864 &	0.778	&	0.830	\\
Virus      &	0.885	& 0.939	& 0.971 &   0.819 &	0.719	&	0.809	\\
Worm       &	0.978	& 0.899	& 0.996 &   0.922 &	0.721	&	0.921	\\

\midrule
\textbf{Average}       &	\textbf{0.967} &	 \textbf{0.920}  &	\textbf{0.9907} &      \textbf{0.848} & \textbf{0.704} &	\textbf{0.865} \\
\bottomrule
\end{tabular}
\end{table}

\captionsetup[subtable]{justification=centering}
\begin{table*}[!htb]
	\centering
	\setlength{\tabcolsep}{1.4pt}
    \caption{Top-10 malware families with the lowest classification accuracy}
	\label{tbl:bestAndWorst_altogether}
    \begin{subtable}{.21\textwidth}
      \centering
		\footnotesize
		\caption{Binary - static}
		\label{tbl:binary_static_bestAndWorst}
		\begin{tabular}{llcc}
		\toprule
			\multirow{2}{*}{\textbf{Family}} & \multirow{2}{*}{\textbf{Class}} &  \textbf{Avg} &  \textbf{\%} \\
			&&  \textbf{Recall} &  \textbf{packed} \\
		\midrule
		pioneer             &       virus &         0.40 &       6\% \\
		asparnet            &    grayware &         0.41 &       5\% \\
		systweak            &    grayware &         0.46 &      19\% \\
		shopper             &    grayware &         0.50 &       1\% \\
		sality              &       virus &         0.52 &       4\% \\
		vitro               &       virus &         0.55 &       3\% \\
		installcore         &    grayware &         0.60 &      10\% \\
		slugin              &       virus &         0.60 &       4\% \\
		elex                &      adware &         0.60 &       9\% \\
		passview            &    grayware &         0.62 &      35\% \\
		\bottomrule
		\end{tabular}
    \end{subtable}
    \begin{subtable}{.21\textwidth}
      \centering
		\footnotesize
		\caption{Family - static}
		\label{tbl:multiclass_static_bestAndWorst}
		\begin{tabular}{llcc}
		\toprule
			\multirow{2}{*}{\textbf{Family}} & \multirow{2}{*}{\textbf{Class}} &  \textbf{Avg} &  \textbf{\%} \\
			&&  \textbf{F1} &  \textbf{packed} \\
		\midrule
		zpevdo              &    grayware &         0.15 &      15\%  \\
		vitro               &       virus &         0.24 &      3\% \\
		uwamson             &    grayware &         0.25 &      15\%  \\
		gendal              &    grayware &         0.28 &      62\%  \\
		dumpex              &    grayware &         0.29 &      40\%  \\
		alman               &       virus &         0.29 &      11\%  \\
		sality              &       virus &         0.33 &      4\% \\
		pasta               &    grayware &         0.34 &      28\%  \\
		cobra               &    grayware &         0.38 &      60\%  \\
		copidmbe            &       virus &         0.39 &      9\% \\
		\bottomrule
		\end{tabular}
    \end{subtable} 
    \begin{subtable}{.27\textwidth}
		\centering
		\footnotesize
		\caption{Binary - dynamic}
		\label{tbl:binary_dyn_bestAndWorst}
		\begin{tabular}{llccc}
		\toprule
			\multirow{2}{*}{\textbf{Family}} & \multirow{2}{*}{\textbf{Class}} & \textbf{Avg} &  \textbf{\%} & \multirow{2}{*}{\textbf{FMR}} \\
			&&  \textbf{Recall} &  \textbf{packed} & \\
		\midrule
		tasker   		&       grayware	   	&  	0.0 	& 11\%  & 0.77	\\
		malex           &       downloader		&   0.0		& 1\%	& 0.77 	\\
		rostpay      	&       grayware    	&   0.0		& 96\%	& 0.76	\\
		constructor     &       grayware      	&   0.0		& 13\%	& 0.78	\\
		atcpa           &       virus     		&   0.0		& 0\%	& 0.78	\\
		mocrt          	&       spyware  		&  	0.0		& 73\%	& 0.80	\\
		mokes           &       backdoor  		&   0.0		& 1\%	& 0.65	\\
		bingoml         &       grayware  		&   0.0		& 22\%	& 0.72  \\
		safebytes       &       grayware  		&   0.0		& 99\%	& 0.81	\\
		trymedia        &       adware  		&   0.0		& 73\%	& 0.70	\\
		\bottomrule
		\end{tabular}
    \end{subtable}
    \begin{subtable}{.27\textwidth}
		\centering
		\footnotesize
		\caption{Family - dynamic}
		\label{tbl:multiclass_dyn_bestAndWorst}
		\begin{tabular}{llccc}
		\toprule
			\multirow{2}{*}{\textbf{Family}} & \multirow{2}{*}{\textbf{Class}} & \textbf{Avg} &  \textbf{\%} & \multirow{2}{*}{\textbf{FMR}} \\
			&&  \textbf{F1} &  \textbf{packed} & \\

		\midrule
		bancos      	&  spyware      &   0.0  	& 44\%	& 0.76	\\
		kovter        	&  grayware     &   0.0 	& 0\%	& 0.78	\\
		safebytes     	&  grayware     &   0.0 	& 99\%	& 0.80	\\
		winner      	&  grayware     &   0.0 	& 0\%	& 0.80	\\
		umbra         	&  downloader   &   0.0 	& 0\%	& 0.80	\\
		ulise        	&  grayware     &   0.0 	& 2\%	& 0.80	\\
		contenedor      &  virus    	&   0.0  	& 0\% 	& 0.80  \\
		cobra     		&  grayware 	&   0.0 	& 60\% 	& 0.79	\\
		kuaizip         &  adware   	&   0.0 	& 1\%	& 0.80	\\
		zpevdo          &  grayware 	&   0.0 	& 15\%	& 0.77	\\	
		\bottomrule
		\end{tabular}
    \end{subtable} 
\end{table*}

This section analyzes which malware classes and families pose
a greater challenge for classifiers based on static and dynamic features.
\revision{Note that our multi-class classification models are for families. 
We only use here the coarser malware class 
(e.g., virus, worm) to draw conclusions on similar families.}

Table~\ref{tbl:overallResultsGrouped} shows 
{Recall} and F1-scores for each malware class in binary and family classification respectively. 
In binary classification, the recall value is defined as the number of correctly classified samples in the class
over the total number of samples in the class.
The numbers differ from those in Table~\ref{tbl:overallResults} because 
Table~\ref{tbl:overallResultsGrouped} only considers the 
classification results of malware samples, 
while Table~\ref{tbl:overallResults} covers the classification of both 
goodware and malware samples (thus taking also false positives into account).

As we can see, the recall and F1 score are not uniform across all classes 
and can widely vary depending on the task and the features used. 
Static features are considerably better at detecting
downloaders, dialers, and worms. In contrast, dynamic features perform better on rogueware, miner, and ransomware. 

These results are confirmed also if we look at individual families. We show
in Table~\ref{tbl:bestAndWorst_altogether} the 10 families with the lowest accuracy in both classification tasks using static and dynamic features. 
For instance, among the 10 malware families for which the static classifier
makes more mistakes, we count four 
\emph{viruses} (i.e., file infectors) and six \emph{grayware}.
This is even more remarkable if we consider the fact that there are only
40 families of Viruses in our entire dataset.
The fact that \emph{viruses} typically append their code to benign files 
results in a wide variation in terms of static features among samples of the same family, 
and this can explain why it is hard for a 
static classifier to differentiate them from \emph{goodware} and from other
families. 
Similarly, \emph{grayware} is defined as undesirable code, which
is not outright malicious per se, therefore making it difficult to find a 
clear boundary to isolate these families. In the worst 10 families using dynamic features,  
we can observe a similar pattern: grayware and viruses dominate the list. 
Besides, adware and spyware are also among the worst families. Malware samples in each of the classes have similar behaviors.

\summary{6}{Models employing static features find it more difficult to classify
\emph{grayware} and \emph{viruses}. 
Dynamic features can identify ransomware, spyware, and adware as malware, but they
have great difficulty in properly identifying their families, probably due to 
very similar runtime behaviors of different families in these classes.}

\begin{table}[t]
\setlength\tabcolsep{2.2pt}
\centering
\footnotesize
\caption{Feature class importance using MDI score.}
\label{tbl:binary_and_multiclass_FeatImportance}
\begin{tabular}{l|rrr|rrr}
\toprule
	\multirow{2}{*}{\textbf{Feature Class}} &  \multicolumn{3}{c}{\textbf{Binary
	classification}} & \multicolumn{3}{c}{\textbf{Family classification}} \\
 	&  \textbf{Comb.} & \textbf{Static} & \textbf{Dyn.} &  \textbf{Comb.} & \textbf{Static} & \textbf{Dyn.} \\
\midrule
s-bytegrams      &  40.88 & 51.38 &     - &  38.60 & 41.67 &     - \\
d-registry       &  17.19 &     - & 25.00 &   0.51 &     - &  0.60 \\
s-opcodegrams    &  13.44 & 21.08 &     - &  23.48 & 20.87 &     - \\
s-strings        &   9.09 & 15.27 &     - &  17.62 & 19.27 &     - \\
d-file           &   7.74 &     - & 29.70 &   3.16 &     - & 56.20 \\
s-sections       &   3.05 &  6.73 &     - &   5.62 &  6.48 &     - \\
s-imports        &   2.48 &  4.17 &     - &   7.87 &  9.30 &     - \\
d-thread         &   2.06 &     - &  7.34 &   0.16 &     - &  5.26 \\
d-network        &   1.51 &     - &  3.50 &   0.35 &     - &  3.70 \\
d-process        &   1.47 &     - & 32.90 &   0.87 &     - & 30.70 \\
s-headers        &   0.34 &  0.72 &     - &   0.73 &  0.96 &     - \\
d-mutex          &   0.25 &     - &  0.16 &   0.03 &     - &  1.19 \\
d-service        &   0.19 &     - &  1.40 &   0.07 &     - &  2.39 \\
s-dll            &   0.17 &  0.28 &     - &   0.52 &  0.57 &     - \\
s-file           &   0.13 &  0.35 &     - &   0.39 &  0.87 &     - \\
\bottomrule
\end{tabular}
\end{table}

\subsection{Feature Class Importance}
\label{sec:eval-features} 

This section examines the importance of the 
static and dynamic features for binary and family classification
using a Random Forest classifier.
We measure feature importance 
using the average Mean Decrease Impurity (MDI) score. 
In a tree-based classifier, 
the MDI score of a feature captures how often the feature was used 
in the tree.
The more a feature is used, 
the more important it is to distinguish different classes. 
For feature classes,
we average the MDI Score across all the features belonging 
to the same feature class and over all the trees in the Random Forest model. 

\paragraph{Feature classes.}
Table~\ref{tbl:binary_and_multiclass_FeatImportance} 
summarizes the feature class importance.
Overall, static features are ranked higher than dynamic features, 
especially for family classification.
This matches results in Section~\ref{sec:classification-results} 
where dynamic features provide marginal improvements over static features.
This observation is in line with recent findings that
although humans prefer dynamic features, 
ML algorithms rely more on the \emph{always present} 
static features~\cite{AonzoUsenix2022}. 

The most contributing static feature classes for both classification tasks 
are \emph{s-bytegrams}, \emph{s-opcodegrams}, and \emph{s-strings}. 
This confirms what was previously observed in the literature, 
with raw and opcode ngrams dominating over other 
static features~\cite{aghakhani2020malware}. 
On the other hand, the most contributing dynamic feature classes for both classification tasks are \emph{d-file} and \emph{d-process}. 
It is interesting to note that even expert human analysts used widely file and process operations to identify malicious behaviours~\cite{AonzoUsenix2022}.
 
In our dataset, over 50\% of the malware samples contain missing features 
values in the \emph{d-network} and \emph{d-service} feature classes, 
thus missing feature values is likely the reason for their low importance.
We evaluate this in Section~\ref{sec:eval-missing}.
%
%
It is interesting that \emph{d-registry} ranks second 
for binary classification, but only 10th for family classification.
This means that registry operations are useful to differentiate 
malware from goodware, but they do not provide enough diversity to 
separate different malware families. 
This likely happens because multiple malware families operate on the same 
registry keys such as those related to achieving persistence (e.g., auto-start) 
and those that disable OS security features. 
In contrast, goodware does not need to operate on those keys.
%

\paragraph{Individual features.}
\revision{
The most contributing static feature classes are
\emph{s-bytegrams} and \emph{s-opcodegrams}, 
but their individual features are hard to interpret. 
For binary classification, 
the top 10 \emph{s-strings} features capture 5 API names 
(\emph{exit}, \emph{CreateThread}, \emph{cexit}, \emph{CopyFileA}, \emph{WinExec}), 
one section name (\emph{.idata}), 
one module name (\emph{MSVCRT.dll}), 
a string possibly related to the .NET runtime 
(\emph{<assemblyIdentity}), and 
two short strings with unclear meaning 
(\emph{:0806}, \emph{L\$ H}).
%
The top \emph{s-sections} features capture section entropy and 
bit 31 in the section characteristics field,
which states if the section can be written to. 
These features are likely related to packing. 
We further examine which static features allow to detect packed malware
in Section~\ref{sec:eval-packing}.
The top \emph{s-imports} features have some overlap with the
top strings (e.g., \emph{exit}, \emph{cexit}), 
but also contain APIs possibly used for evasion
(e.g., \emph{queryperformancecounter}, \emph{getsystemtimeasfiletime}) and 
popular C runtime functions
(e.g., \emph{free}, \emph{calloc}, \emph{malloc}, \emph{fprintf}).
For family classification, the top static individual features differ 
from those for binary classification with no intersection 
between the top 10 \emph{s-strings} and \emph{s-imports} 
for binary and family classification.
For example, the top strings contain 6 API names 
(\emph{WNetOpenEnumA}, \emph{WNetEnumResourceA}, \emph{WNetCloseEnum},
\emph{RegisterServiceProcess}, 
\emph{PathFileExistsA}, \emph{UpdateResourceA}), 
a third-party library name (\emph{StringX}), and 
some short strings 
(\emph{QQQQS3}, \emph{lllll}, \emph{3.91}).
These strings are not highly ranked for binary classification and are possibly 
associated with specific families.
}





\revision{
Among the dynamic features, the most contributing classes are
\emph{d-file} and \emph{d-process}.
In contrast to the static features, 
the top contributing dynamic features largely overlap 
between binary and family classification.
The top process features are 
the number of processes invoking shell commands, and
the number of terminated, opened, and created processes.
The top file features capture the entropy of the files accessed, 
as well as the name of some specific files, such as
\emph{appdata{\textbackslash}local{\textbackslash}temp{\textbackslash} 7zipsfx.000}, 
which likely indicates the executable is an SFX installer.
One difference between binary and family classification is that 
for family classification the number of mutexes created is a top contributor. 
Mutexes are often used by malware creators to avoid re-infecting the same 
host and their number and values are intuitively family-specific.
}

\revision{Overall, the interpretability of individual features can be hard, 
especially for n-grams.
In fact, we argue that one benefit of ML classifiers is that they can select 
the features they consider most useful, which a human may not be able to 
identify based on domain knowledge.
Our data release~\cite{anon_repo} includes the top 
individual features for the different models.}

\summary{5}{
Static features are more important than dynamic features for 
both classification tasks, but especially for family classification. 
Raw and opcode n-grams are the most important feature classes in 
both classification tasks. 
The importance of a feature class may depend on the classification task.
For example, \textit{d-registry} is important to distinguish malware from 
goodware, but is not relevant for family classification. 
}

\subsection{Impact of Packers and Protectors} \label{sec:eval-packing}
This section evaluates whether the presence of off-the-shelf packers and
protectors harms the classification accuracy when considering static features.
\revision{Our dataset comprises real malware collected from a commercial feed, 
so we expect the fraction of packed samples to approximate that in the wild. 
Overall, we identified 
119 unique known packers, including 
highly sophisticated ones like VMProtect and Themida, covering 22\% of the samples
in our dataset.
However, this ratio is certainly a lower bound as packer detection tools may not identify 
custom packers. 
Tables~\ref{tbl:binary_static_bestAndWorst}--\ref{tbl:multiclass_dyn_bestAndWorst}
show that the packing rate largely varies per family: some have
99\% of their samples packed while others have none. 
As explained in Section~\ref{sec:features-static}, 
we did not attempt to unpack samples, 
but follow prior work in extracting static features regardless of whether a file 
is packed or not. The packer information is only used for the analysis of the results.}

We first investigate whether the models overfit the packers
or instead can capture data that allows
them to classify samples correctly.
To answer this question, we first compute the family-wise classification
accuracy for both binary and family classification using static features.  We
then compute the Pearson correlation scores between the family-wise accuracy
scores and the rate of packed samples in each family.  If packing negatively
affects the ability to classify a sample, we would expect lower accuracy for
families where packing is more prevalent.
However, the correlation scores are 0.015 and 0.0001 respectively for binary 
and family classification. To statistically support these results, we run a T-test 
with the null hypothesis being that there is not a significant correlation 
between classification accuracy and packing presence. We respectively obtain
0.51 and 0.98 as p-values that do not allow us to reject the null hypothesis.
Thus, we conclude that there is not a statistically significant correlation 
between the two variables.
%
%
This might seem surprising, as one might expect a high correlation between
packing and misclassification rate at least for models that rely only on static
features. After all, packing was one of the main reasons that led researchers to
introduce malware analysis sandboxes and dynamic analysis.  However, this is a
common misconception. In fact, while packing is very effective at impeding
static \textit{analysis} (i.e., the ability to examine a sample and statically
derive its behavior), other works~\cite{aghakhani2020malware} have shown that
common packers leave certain areas of the binary untouched, thus having a
limited effect on the ability of a ML \textit{classifier} to identify a sample.
\revision{While our static models seem capable to detect samples 
protected with off-the-shelf packers, 
newer protectors can be designed to specifically target static models. 
Also, it is possible that some of the hard-to-detect families use 
(undetected) custom packers that indeed hamper the detection.}

\begin{table}[t]
\setlength\tabcolsep{2pt}
\centering
\footnotesize
\caption{\revision{Feature class importance using MDI score when considering 
all, packed only, and not-packed samples only.}}
\label{tbl:binary_and_multiclass_packed_only_FeatImportance}
\begin{tabular}{l|rrr|rrr}
\toprule
	\multirow{2}{*}{\textbf{Feature Class}} &  \multicolumn{3}{c}{\textbf{Binary
	classification}} & \multicolumn{3}{c}{\textbf{Family classification}} \\
 	&  \textbf{All} & \textbf{Packed} & \textbf{Not-Packed} &  \textbf{All}
	& \textbf{Packed} & \textbf{Not-Packed} \\
\midrule
s-bytegrams      & 51.38 & 62.22 & 49.30 &  41.67 & 53.66 &  38.59 \\
s-opcodegrams    & 21.08 & 8.30  & 22.69 &  20.87 & 9.95  &  25.02 \\
s-strings        & 15.27 & 16.80 & 16.16 &  19.27 & 18.17 &  17.80 \\
s-sections       &  6.73 & 7.50  & 6.29  &   6.48 & 9.39  &  10.17 \\
s-imports        &  4.17 & 2.29  & 4.35  &   9.30 & 5.32  &  6.09  \\
s-headers        &  0.72 & 1.42  & 0.63  &   0.96 & 1.30  &  1.17  \\
s-dll            &  0.28 & 1.06  & 0.21  &   0.57 & 0.91  &  0.78  \\
s-file           &  0.35 & 0.40  & 0.36  &   0.87 & 1.29  &  0.36  \\
\bottomrule
\end{tabular}
\end{table}

\revision{To understand which static features are more effective at  
identifying packed malware, 
we compute the importance of the feature classes separately for two sets: packed samples on one side 
and unprotected (i.e., not packed) samples on the other.
Table~\ref{tbl:binary_and_multiclass_packed_only_FeatImportance} 
summarizes the results for both binary and family classification. 
The \emph{All} column captures the feature importance for all samples 
(regardless of packing) and thus matches the values already reported in 
Table~\ref{tbl:binary_and_multiclass_FeatImportance}. 
The results show that for both binary and family classification of packed samples, 
the relative importance of \emph{s-bytegrams} increases significantly 
(compared to all samples) and 
there are also relevant increases in the importance of 
\emph{s-sections}, \emph{s-headers}, and \emph{s-dll}. 
On the other hand, the relative importance of \emph{s-opcodegrams} and
\emph{s-imports} is greatly reduced.}

\revision{This is likely due to the fact that much of the code in packed samples is compressed or 
encrypted, reducing the amount of useful opcodes that can be extracted statically to those in the unpacking routine. 
On the other hand, raw bytegrams are still able to capture distinctive 
sequences of bytes, which may act like signatures for the packed samples. 
Those sequences can be extracted from parts of the executable that 
are not code (e.g., PE header and data sections).
The classifier focusing on those parts for packed samples would also 
explain the increased importance of \emph{s-sections}, \emph{s-headers}, 
and \emph{s-strings}.
In addition, some packers use weak encryption schemes based on XOR operations with a fixed 
key, which may make distinctive byte sequences in the unpacked code to still 
be distinctive (in their encrypted form) in the packed executable.
The decrease in importance for \emph{s-imports} is likely linked to 
packers obfuscating the import table. 
Finally, most packers leave a very reduced import table that 
tends to use the same Windows libraries, 
which could explain the slight increase for \emph{s-dll}.
}



\summary{3}{ Packed or protected samples (with off-the-shelf tools) do not
significantly correlate with their classification accuracy using static
features.  This means that although these technologies function well to deter
static analysis (in particular reverse engineering), 
they do not significantly affect
ML classifiers, \revision{which are still able to successfully identify byte-level signatures}.}


\subsection{Impact of Missing Dynamic Feature Values}
\label{sec:eval-missing}

Some possible explanations for the worse results of dynamic features 
compared to static features 
are that a sandbox may fail to stimulate samples adequately 
to cause them to `detonate`, or that samples may not work properly due 
to missing local or remote components. 
As a result, the classifier might need to take a decision based on a 
partial view of the malware runtime behavior. 

%


We computed the Pearson correlation coefficient between the family-wise recall
of binary classification and the FMR to study the link between the two. Interestingly, 
the correlation is not statistically significant for the binary
classification task (pearson -0.1 and p-value 0.11). However, there is a
clear negative correlation (-0.43, p-value of $7.61*10^{-16}$) for the
family classification task. In this case, as the fraction of samples with
missing feature values for a family increases, its classification accuracy
decreases. 
This is also confirmed by looking at the malware families that are the most difficult
to classify with dynamic features, i.e., those for which the classifier
has the lower accuracy (see Tables~\ref{tbl:binary_dyn_bestAndWorst} and~\ref{tbl:multiclass_dyn_bestAndWorst} in Section.\ref{sec:best_and_worst}). Among the top-10 all have an FMR $>$ 65\%.

This outcome demonstrates that the ML classifier
might still be able to identify signs of malicious behavior in incomplete
dynamic analysis reports, but more feature values are needed to precisely
distinguish among different families (in particular for those, like
downloaders, that might have similar behavioral profiles).
In addition, binary classification is also affected by the quality of the 
behaviors collected from benign samples, while
family classification accuracy is solely 
associated with the feature completeness of malware samples in each family.




\summary{2}{
Globally, a statistically significant inverse correlation in
the family classification task between the family-wise classification
accuracy using dynamic features and the amount of missing dynamic feature
values exist. 
The correlation is instead not significant for the binary classification task.
}


\subsection{Impact of Ground Truth Confidence}
\label{sec:eval-avclass}

To assign a family to a sample 
\avclass computes a list of (tag, confidence) pairs,
e.g., (FAM:sality, 5), (CLASS:virus, 4), (FAM:zpevdo, 1).
Then, it selects as family the highest confidence tag 
that is either a family in its taxonomy or 
an unknown tag not in its taxonomy.
%
The confidence score roughly represents the number of AV engines that 
assign a tag to the sample, after accounting for aliases and 
discounting groups of AV engines that copy their labels.
%
This section examines whether the \avclass confidence score 
for the selected family impacts the classification accuracy. 

To examine this issue, we first compute the confidence score for each family. 
For each sample, we obtain a normalized confidence in the [0,1] range
by dividing the confidence score of the assigned family 
over the sum of the confidence scores for 
all family and unknown tags for the sample.
In the case above, this step returns 0.83 as the \emph{FAM:sality} confidence 
was 5, but \emph{FAM:zpevdo} also appeared in the output.
Then, we average the normalized confidence factor across all samples 
in the family to produce a family confidence score.

Next, we compute the correlation between the 
family-wise classification accuracy and the family confidence score.
The hypothesis is that higher family confidence scores correlate with 
higher family classification accuracy, 
i.e., the more agreement AV engines have when tagging the sample, 
the easier it should be to classify the sample. 
The Pearson correlation coefficient is  
0.083 for static features (p-value 0.03) 
and 0.062 for dynamic features (p-value 0.01). 
The correlation is positive but extremely small. Thus,
we can conclude that 
poor family classification is not influenced by 
a low \avclass confidence score and the result is statistically
significant.
This is further confirmed by examining the 10 families with the lowest 
classification accuracy using either static-only or dynamic-only features (Table~\ref{tbl:binary_static_bestAndWorst} and Table.\ref{tbl:binary_dyn_bestAndWorst}). 
Of those 20 families, all have a confidence score above 0.5 and 
15 have a confidence score above 0.8.
This suggests that even when the AV engines do not fully agree on the name 
of a sample, the majority vote likely selects the correct family, 
which provides further confidence on our \avclass-based 
ground truth generation approach. 

\summary{4}{The accuracy of family classification is not correlated 
    with the \avclass confidence score, which captures the agreement
    between different AV vendors on the family name of a sample.
    This observation supports that AVclass2 is a valid tool for getting ground 
    truth when it is necessary to obtain the family name of malware.
}

\subsection{Impact of Training Dataset Construction}
\label{sec:dataset_construction} 

This section evaluates the effect of the construction of the training dataset
on classification accuracy. We specifically investigate the impact of the size
of the training dataset, the variety of malware families represented, and the
uniformity of the sample-family selection.
To the best of our knowledge, the question of how diversity in terms of
families impact binary classification has not been studied before. 

To study this aspect we plot a number of heatmaps. In each experiment,
as described in Section.\ref{sec:methodology},
we reserved randomly 20 samples in each 
family for validation (e.g., hyper-parameter tuning) and we choose
$p$ samples from the remaining 80 samples and use them for training and testing.
To study the impact of number of available samples, we vary $p$ from 50 to 80. 
To study instead the impact of the number of different families in the dataset, we progressively
vary the number of
families involved in both binary and family classification from 70 to
670. For each combination of number of families and number of samples per family, we
conduct a 10-fold cross validation test and report the averaged F1 score in
the corresponding cell of each heatmap.

\begin{figure}[t] \centering
\includegraphics[width=0.8\columnwidth]{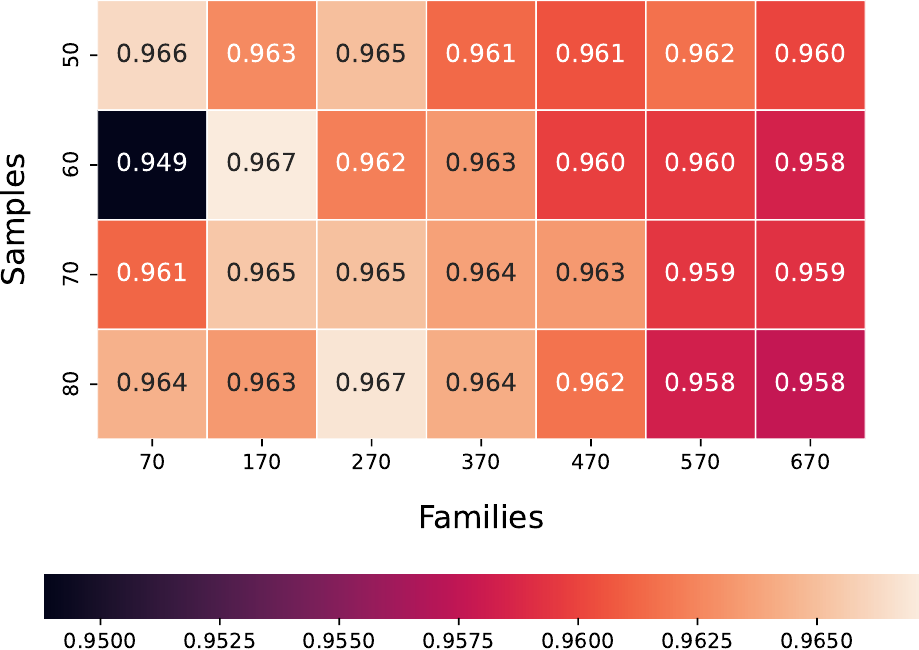}
\caption{F1 score heatmap for \textbf{binary classification} using static
model.} \label{fig:binary_static_f1_heatmap} 
\end{figure}

\begin{figure}[t] \centering
\includegraphics[width=0.8\columnwidth]{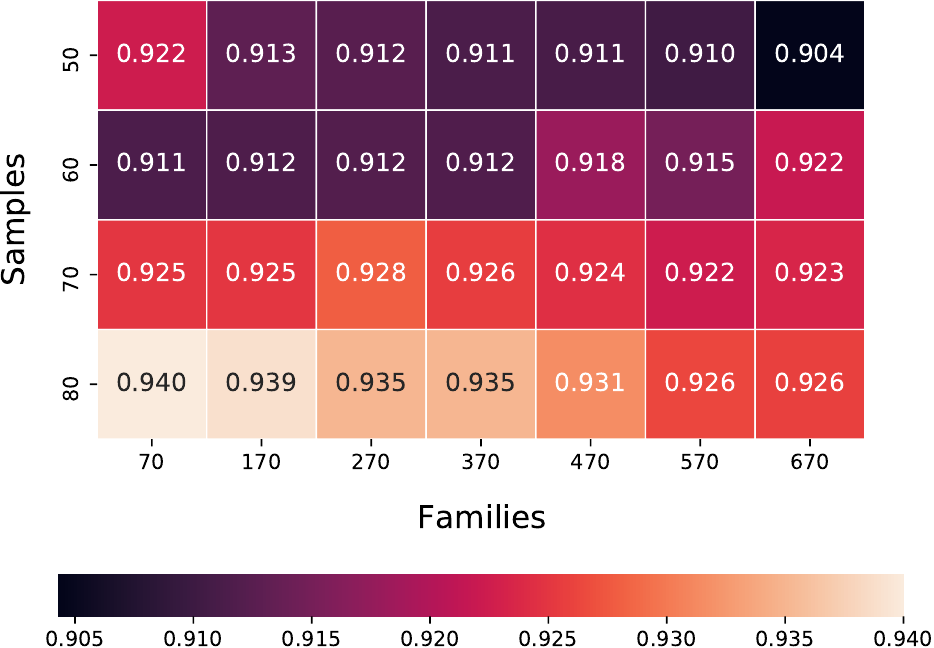} \caption{F1
score heatmap for \textbf{binary classification} using dynamic model.}
\label{fig:binary_dynamic_f1_heatmap} 
\end{figure}

\begin{figure}[t]
\centering
  \includegraphics[width=0.6\columnwidth]{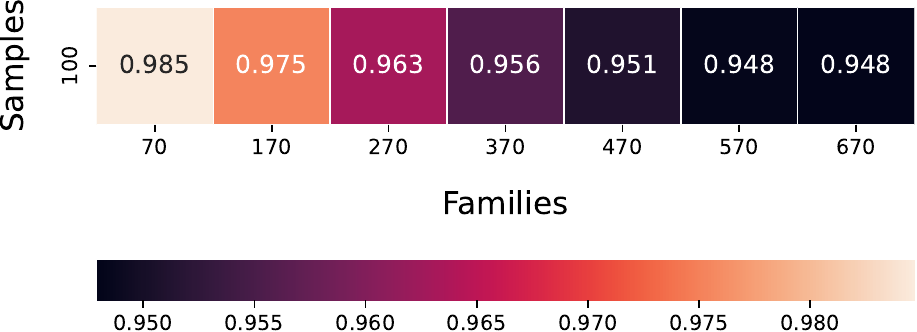}
  \caption{F1-score heatmap for \textbf{binary classification} with combined model.}
\label{fig:binary_combined_f1_heatmap}
\end{figure}

\begin{figure}[t]
\centering
  \includegraphics[width=0.8\columnwidth]{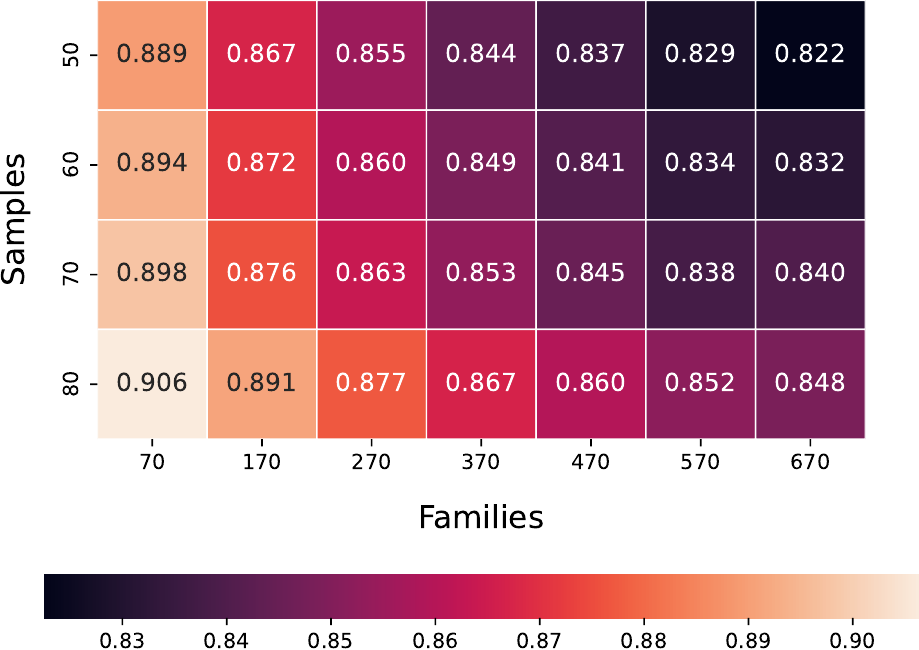}
  \caption{F1 score heatmap for \textbf{family classification} using Random Forest on static analysis features.}
\label{fig:multiclass_static_f1_heatmap}
\end{figure}
\begin{figure}[t]
\centering
  \includegraphics[width=0.8\columnwidth]{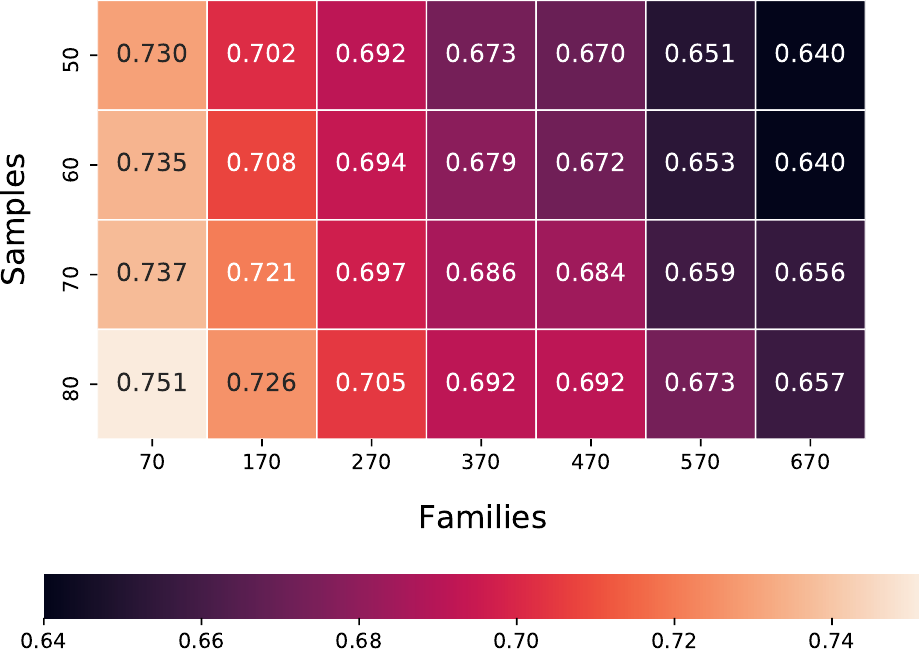}
  \caption{F1 score heatmap for \textbf{family classification} using Random Forest on dynamic analysis features}
\label{fig:multiclass_dynamic_f1_heatmap}
\end{figure}
\begin{figure}[t]
\centering
  \includegraphics[width=0.6\columnwidth]{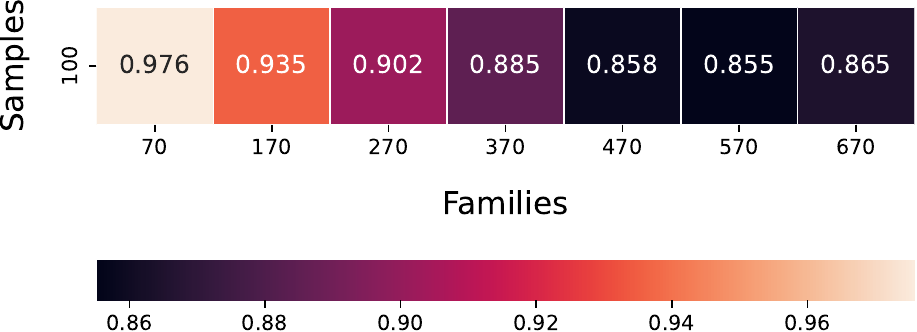}
  \caption{F1-score heatmap for family classification when combining features derived from static and dynamic analysis}
\label{fig:multiclass_combined_f1_heatmap}
\end{figure}

Figure~\ref{fig:binary_static_f1_heatmap} and 
Figure~\ref{fig:binary_dynamic_f1_heatmap}
present heatmaps of the F1 score for binary classification, 
using static features and dynamic features respectively.
%
Figure~\ref{fig:binary_combined_f1_heatmap} shows the heatmap for the 
combined model, for brevity only showing the variation with the 
number of families.
Figure~\ref{fig:multiclass_static_f1_heatmap},
Figure~\ref{fig:multiclass_dynamic_f1_heatmap}, and
Figure~\ref{fig:multiclass_combined_f1_heatmap} 
are similar but for the family classification task.

Overall, the results indicate that
as the number of samples per family 
increases, the classification accuracy also increases.
The exception is for the binary classification using static features, 
where increasing the samples per family may cause a decrease in overall 
accuracy. 
For example, when using 50 samples for each of the 670 families 
the F1 score is 0.960, but when using 80 samples it 
slightly decreases to 0.958. 
However, the trend is different if we consider more families. 
We consider these very small changes as fluctuations due to the randomness of the sample selection process.
%

With respect to family diversity, 
the results confirm that the more families in the training dataset 
the more difficult their classification is.
As expected, the decrease in classification accuracy is more marked for the 
family classification task, 
where intuitively the higher the number of classes the more difficult the 
classification becomes.
The decrease is also more marked for the dynamic features than for the 
static ones, likely due to their lower discriminatory power 
as discussed in Section~\ref{sec:classification-results}.

\paragraph{Non-uniform sampling.}
We also evaluate the impact of a non-uniform downsampling strategy
for binary classification. 
For this purpose, we mimic the distribution of the recently-proposed MOTIF dataset~\cite{motif}, 
which contain 3,095 PE malware samples from 454 families with an unbalanced 
distribution 
(e.g., the median is 3 samples per family and 
29\% of families have a single sample). 
We create a new dataset by applying the MOTIF distribution 
to $M_B$.
This new MOTIF-like dataset comprises 4,821 samples from all 
670 families with the following distribution:
29\% of the families are singletons, 
41\% have 2-5 samples,
12\% 6-10,
10\% 11-20,
4\% 21-30,
2\% 31-40,
1\% 41-50,
and 1\% has over 50 samples (up to 100). 


We use this to compare two sampling approaches: 
the \emph{uniform} approach (which is the one we adopted so far in the paper) where we keep a balanced number
of samples for each family, versus a \emph{nonUniform} approach, where we consider a real-world case in which
the number of available samples varies from one family to another, as captured by the MOTIF-like dataset.
Table~\ref{tbl:binary_accuracy} shows the results for both approaches and 
different feature sets. 
We could not identify any significant difference between the two approaches,
thus suggesting that training a classifier with a non-uniform amount of samples
does not significantly impact its performance, under the important assumption that
the testing dataset also follows the same distribution.


\summary{7}{
Increasing the number of malware families in the training set
makes the classification more complex and generally results in lower accuracy. 
While not surprising, this is very important because previous studies were often 
performed on only a few dozens of families, with the risk of reporting inflated
results that do not generalize to larger and more realistic datasets.

Increasing the number of samples per family can help to increase the 
classification accuracy, in particular for models based on dynamic analysis.
Finally, the choice between a non-uniform and a uniform downsampling strategy 
does not significantly affect the binary classification accuracy.
}
\begin{table}[t]
\centering
\small
\caption{Impact of uniform and non-uniform sample selection in training dataset.}
\label{tbl:binary_accuracy}
\begin{tabular}{lrrrr}
\toprule
\textbf{Model} &  \textbf{Prec.} &    \textbf{Recall} &  \textbf{F1} &  \textbf{Acc.} \\
\midrule
binary-static-uniform   &   0.956 &  0.957 & 0.957  & 0.957  \\
binary-dynamic-uniform & 0.962 & 0.892 & 0.926 & 0.929\\ 
binary-combined-uniform & 0.963 & 0.934  &  0.948 & 0.948 \\
\hline
binary-static-nonUniform  & 0.961 & 0.960 & 0.961 & 0.960 \\ 
binary-dynamic-nonUniform & 0.959 & 0.886 & 0.921 & 0.924 	\\ 
binary-combined-nonUniform& 0.955 & 0.927 & 0.940 & 0.927 	\\ 
\bottomrule
\end{tabular}
\end{table}

\subsection{Model Generalization}
\label{sec:singleton_and_unseen} 

\begin{table}[t]
\caption{Binary classification accuracy on singletons and unseen families 
datasets.}
\label{tbl:binary_singletons_unseen}
\small
\begin{tabular}{lrrrr}
\toprule
\textbf{Model} & \textbf{Singletons} & \textbf{Unseen} \\
\midrule
binary-static-uniform  & 0.943 & 0.815 \\
binary-dynamic-uniform & 0.805 & 0.898 \\
binary-combined-uniform& 0.985 & 0.908\\
\hline
binary-static-nonuniform  & 0.810 &  0.653 \\ 
binary-dynamic-nonuniform & 0.328  &  0.855\\ 
binary-combined-nonuniform& 0.758  & 0.637 	\\ 
\bottomrule
\end{tabular}
\end{table}

\begin{figure}[h] \centering
	\includegraphics[width=0.8\columnwidth]{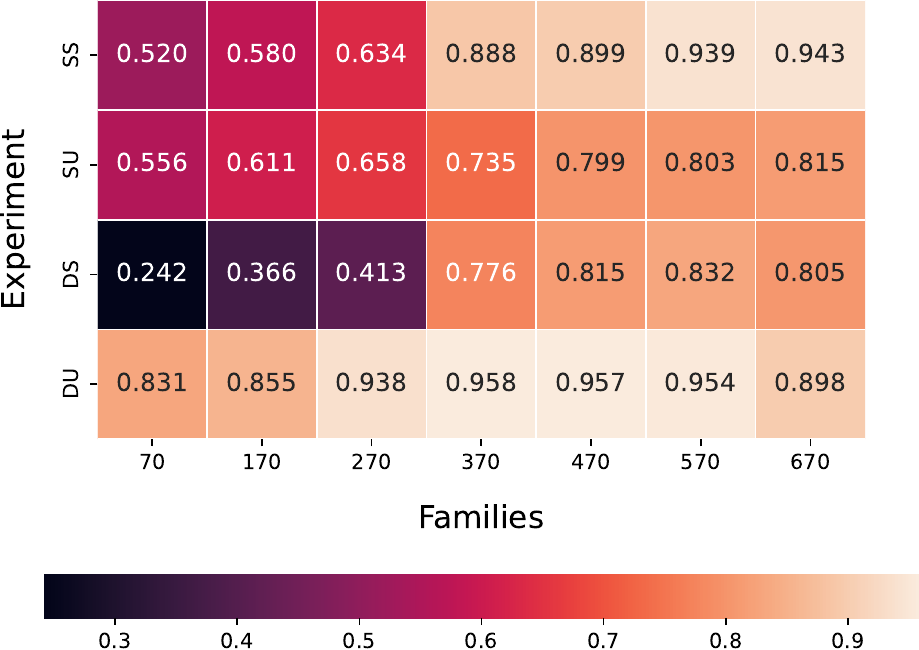}
	\caption{Binary classification accuracy on singletons and unseen families
	of the uniform dynamic and static models. (SS: Static
Singleton. SU: Static Unseen. D is for Dynamic)}
\label{fig:binary_singletons_unseen_heatmap}
\end{figure}


In this section, we test how well our models for binary and family
classification generalize on unseen data. To this extent, we validate the performance of
the previously-trained models on the singleton and unseen 
datasets introduced in Section~\ref{sec:maltest}, which include new families and have 
different distributions from the training data.
This scenario is known as the "out-of-distribution" (OOD)
test~\cite{Liu2020nips}, where training and testing data have different
distributions in the feature space. 
The distribution gap between the training and testing data has been frequently
witnessed in malware analysis~\cite{Jordaney2017usenix}, as
malware families evolve rapidly over time. 
Theoretically, one should expect the performance of a ML model 
to drop drastically in this more realistic scenario, as OOD samples directly violate the IID assumption of ML techniques. 


\paragraph{Binary Classification.} Table~\ref{tbl:binary_singletons_unseen}
summarizes the binary classification results over the singletons and unseen
families using the static, dynamic, and joint feature pool.
"Uniform" and "non-uniform" in the table denote training with the 670 families
with uniformly and non-uniform dataset construction methods (\S~\ref{sec:models})
The empirical measurements shown in Table~\ref{tbl:binary_singletons_unseen} can
be summarized around three main observations. 

First, the accuracy of binary classification using only
static or dynamic features deteriorates significantly over singleton and unseen family files. 
Using the combined feature set, the binary classification accuracy with the
uniform setting augments over the singleton samples, whereas it deteriorates over the unseen families. 
In the non-uniform setting, we can observe the same
tendency of accuracy drop over the OOD samples. The observations echo closely to the
out-of-distribution challenge of machine learning raised in
\cite{Liu2020nips}. 

Second, the accuracy deterioration over the
out-of-distribution samples is more significant in the non-uniform setting of training than
that in the uniform setting, regardless of the used features. 

This is different from the results of the
in-distribution evaluation in Table~\ref{tbl:binary_accuracy}, where we observe no major
difference in accuracy between the uniform and non-uniform settings. These
results show an important point: classifiers built on very unbalanced datasets may perform equally well
when tested on samples with the same unbalanced distribution, but generalize more poorly to 
other testing datasets, likely because many families were underrepresented in the training
and thus the model failed to properly capture them.


Third, we can notice that static features generalize poorly to unseen families, while dynamic
features perform better in this scenario. This is due to the nature of the features
themselves: static information can precisely pinpoint only known samples, while
dynamic behavior can better generalize also to unknown ones.
Thus, compared to static features, dynamic features may provide more
rich information to capture new types of malicious behaviors that never
appear in the training phase.  


We investigate this aspect in more detail by varying the number of families 
we used for training. In Figure~\ref{fig:binary_singletons_unseen_heatmap}, we can see that 
dynamic features perform
poorly when the number of malware families for training is low (as there was not enough
example of behaviors to learn from) but, with a sufficient number of families, they
offer better classification results than static features. 
Dynamic features usually have a high dimensional
and highly sparse feature representation. For example, some files or
processes only appear a few times in the training set for specific
malware families. A smaller number of families may aggravate the
curse of dimensionality, which results in an overfitting of the classifier.
Furthermore, we can observe the classification
accuracy over unseen samples improves as the number of families increases,
regardless of the features used in the test.

\begin{figure}[t] \centering
\includegraphics[width=0.8\columnwidth]{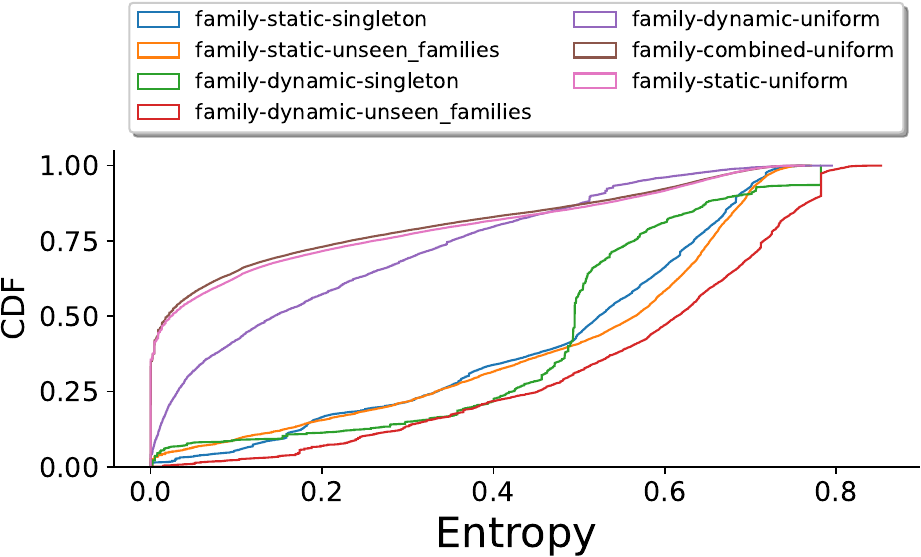}
\caption{Entropy distribution comparison} \label{fig:entropyComparison}
\end{figure}

\paragraph{Family Classification.}
So far, we only tested the generalization of our models in a binary classification
scenario. We now apply our family classifier
trained using the 670 families over the singleton and unseen families as another
out-of-distribution test scenario. Achieving high or low classification accuracy
over these out-of-distribution samples is not interesting, as most of these
samples share no common families as the training data and we don't have the
ground truth family labels for these samples. Thus, the purpose of organizing this
test is only to study how the uncertainty level of the family classifier changes
over the out-of-distribution malware samples. 


To measure the uncertainty difference, we define the Relative Entropy Score
(\textit{RES}) of the classifier's output as $\frac {\sum_{k=1}^{C} p_{k}\log{p_{k}}} {T}$, 
where $T={
\sum_{k=1}^{C} 1/C\log{1/C}}$ and $C$ is the number of the
families covered by the training data building the classifier. In this
experiment, $C$ is therefore set to 670.  For an input sample, the output of the family classifier is a 670-dimensional
probability-valued vector $\{p_{k}\}$ (k=1,2,3,...,C=670). Each $p_{k}$ gives
the probabilistic confidence that the sample belongs to the corresponding
family.  By definition, the numerator $\sum_{k=1}^{C} p_{k}\log{p_{k}}$ provides
the entropy of the classifier's output. The denominator $\sum_{k=1}^{C}
1/C\log{1/C}$ denotes the maximum entropy that the classifier's classification
output may have. 
As a result, the magnitude of \textit{RES} is strictly normalized between 0 and 1. Higher/Lower
\textit{RES} denotes that the classifier shows higher/lower uncertainty level
over the classification output.


In Figure~\ref{fig:entropyComparison}, we demonstrate the empirical cumulative
distribution function (CDF) of RES-based uncertainty distribution of the family
classifier's output on the testing malware samples of the 670 families
(in-distribution samples) and those belonging to singleton / unseen families. 
Consistently with theoretical studies~\cite{Liu2020nips}, we can find that the uncertainty level of the family classification output over the singleton and malware samples of previously unseen families increases significantly, compared to those derived with the testing samples sharing the same families of the training data.

\summary{8}{Our experiments confirm a significant performance drop in binary classification
    over out-of-distribution samples, both in the case of singleton and unseen
    families. At the same time, the confidence of the ML-based
    classifier decreases significantly over these out-of-distribution
    samples. This implies that ML-based models tend to be less certain over
    malware samples drifted from the training samples. 
    Our results also show that models trained on a very unbalanced dataset generalize
    more poorly, and that dynamic features generalize better than static over
    new families. Overall, as the distribution
    gap between training and testing malware samples is common in
    practice, these results raise concern over the utility of ML-based
    malware classification for real-world scenarios. } 

\section{Related Work}
\label{sec:related}

\begin{table}[t]
\centering
\scriptsize
\caption{Related work on ML-based Detection and family Classification 
of Windows malware {\small(S=Static, D=Dynamic)}}
\begin{tabular}{|l|c|c|c|c|c|r|r|}
\cline{3-8}
\multicolumn{2}{c|}{} &
\multicolumn{2}{c|}{{\bf Goal}} & 
\multicolumn{2}{c|}{{\bf Features}} &
\multicolumn{2}{c|}{{\bf Dataset}} \\
\hline
 \textbf{Work}
 &\textbf{Year}
 &\textbf{D}
 &\textbf{C}
 &\textbf{S}
 &\textbf{D}
 &\textbf{\#}
 &\textbf{Fam.} \\
\hline
Rieck et al.~\cite{rieck2008learning} & 2008 & \N & \Y & \N & \Y & 10k & 14 \\
McBoost~\cite{mcboost} & 2008 & \Y & \N & \Y & \Y* & 5.5k & - \\ 
PE-Miner~\cite{shafiq2009pe} & 2009 & \Y & \N & \Y & \N & 16k & - \\
Nataraj et al.~\cite{nataraj2011comparative} & 2011 & \N & \Y & \Y & \Y & 67k & 561* \\ 
OPEM~\cite{santos2012opem} & 2012 & \Y & \N & \Y & \Y & 1k & - \\
Santos et al.~\cite{santos2013opcode} & 2013 & \Y & \N & \Y & \N & 1k & - \\
Dahl et al.~\cite{dahl2013large} & 2013 & \Y & \N & \N & \Y & 1.8M & - \\
Kancherla et al.~\cite{kancherla2013image} & 2013 & \Y & \N & \Y & \N & 25k & - \\
Saxxe et al.~\cite{saxe2015deep} & 2015 & \Y & \N & \Y & \N & 350k & - \\
Miller et al.~\cite{miller16reviewer} & 2016 & \Y & \N & \Y & \Y & 1.1M & - \\ 
MtNet~\cite{mtnet} & 2016 & \Y & \Y & \N & \Y & 2.8M & 98 \\
MAAR~\cite{maar} & 2017 & \Y & \N & \N & \Y & 3k & - \\
MalConv~\cite{malconv} & 2018 & \Y & \N & \Y & \N & 284k & - \\
EMBER~\cite{anderson2018ember} & 2018 & \Y & \N & \Y & \N & 400k & - \\
Rhode et al.~\cite{rhode2018early} & 2018 & \Y & \N & \N & \Y & 5.1k & - \\
MalDy~\cite{maldy} & 2019 & \Y & \Y & \N & \Y & 20k & 15 \\
NeurLux~\cite{neurlux} & 2019 & \Y & \N & \N & \Y & 34k & - \\
MalInsight~\cite{malinsight} & 2019 & \Y & \Y & \Y & \Y & 3.5k & 5 \\
MalDAE~\cite{han2019maldae} & 2019 & \Y & \N & \Y & \Y & 5.5k & - \\ 
MALDC~\cite{maldc} & 2020 & \Y & \N & \N & \Y & 54k & - \\
IMCFN~\cite{vasan2020imcfn} & 2020 & \Y & \N & \Y & \N & 9.4k & - \\
Zhang et al.~\cite{zhang2020dynamic} & 2020 & \Y & \N & \N & \Y & 27.7k & - \\
Rabadi et al.~\cite{rabadi2020advanced} & 2020 & \Y & \N & \N & \Y & 7.1k & - \\ 
Joyce et al.~\cite{motif} & 2022 & \N & \Y & \Y & \N & 3k & 454 \\
\hline
This work & 2023 & \Y & \Y & \Y & \Y & 67k & 670 \\
\hline
\end{tabular}
\label{tbl:related}
\end{table}




Table~\ref{tbl:related} presents a categorization of previous
works on Windows malware classification, according to their
goal (binary detection or family classification), 
features (static or dynamic), and 
dataset size (both in terms of malware executables and malware families). 
Among the approaches in Table~\ref{tbl:related}, 
the choice of the models varies widely including classical models like 
Support Vector Machine, GradientBoost, and Random Forest,
as well as neural networks. 
Most approaches perform feature extraction, 
e.g. extract n-grams of bytes, opcodes, or system calls, 
but a couple of work directly operate on raw bytes and 
API sequences~\cite{malconv,neurlux}.
 
MalInsight~\cite{malinsight} is the only study so far to provide a comprehensive 
coverage over the choice of features and classification tasks. 
However, their dataset includes only 5 families. 
At the other end of the spectrum,
Nataraj et al.~\cite{nataraj2011comparative} studied only family classification
on an unbalanced dataset with over 500 classes. 
However, the authors consider each full AV label a different class, 
so that number does not correspond to real malware families.
In contrast, our study investigates the factors impacting the performance of 
ML classifiers using a large-scale balanced dataset with 670 families. 

\paragraph{ML challenges and pitfalls.}
In cyber security research, two major challenges are raised in the practices of ML-based analysis. First of all, the issue of missing observations affects the prediction accuracy, e.g., in network intrusion detection~\cite{Pawlicki2021IIDS, TavabiWWW20}. Secondly, most ML models follow a core assumption: the training and test data of a ML model should be drawn identically and independently from the same underlying distribution, i.e. the I.I.D. assumption. However, the I.I.D assumption does not hold in practice. Highly diversified and quickly evolving malware technologies make the implementations and behaviours of malware vary significantly and frequently. New variants of malware arise to exploit novel vulnerabilities and evade the detection of anti-virus services. Once a machine-learning-driven malware classifier is deployed in practical security applications, the fast-changing profiles of malware samples break the I.I.D, assumption and cause the deterioration of the classification accuracy  \cite{transcending2022SP}. 
The design of a robust classifier for frequently drifting malware profiles is still an open problem. 

Arp et al.~\cite{Arp2022Usenix} review the use ML-based classification in
cyber security published over the past 10 years. The study summarizes the
existing issues at the different stages of the ML-based pipelines for cyber
security data analysis. For example, the authors demonstrate that the statistical
bias introduced by training sample sampling and inaccurate class label
tagging may introduce spurious correlations into the ML classifiers.
In addition, employing inappropriate performance metrics ignoring the class
imbalance in the testing phase may lead to incorrect interpretation to the
quality of ML-based predictive analysis. In general, according 
to~\cite{Arp2022Usenix}, the performance metrics of a ML-based analysis
pipeline in cyber security practices should be defined by considering the
characteristics of the security data collected and the requirements raised
in the concerned applications. Otherwise, the pipelines may produce
unrealistic performance and interpretations of security incidents.
In our work, we focus instead on the bottlenecks of ML-based malware classification encountered in practices, which may obstruct the accurate classification of malware. 
For instance, we focus on the
impact of the coverage of malware families for training and we dive into the
potential reasons causing failure of ML-based models over certain malware
samples. We also explore how the classifier behaves over out-of-distribution
malware samples, which is an interesting problem in the practical
deployment of ML-driven pipelines.


\paragraph{Dataset construction.}
In 2015, the Microsoft Malware Classification Challenge
\cite{smith2020mind} was developed as a Kaggle competition to conduct
malware family classification. The corresponding dataset is composed of disassembly and
bytes of 20K Windows malware samples from 9 families. It was released in
the Kaggle competition and has since been used in several studies.
\cite{tesseract,lee2021android} built larger-scale Android malware datasets
for evaluating the performances of ML-driven classification models. More
specifically, \cite{tesseract} evaluates the spatial and temporal bias of
binary classification accuracy over 129,728 Android apps.
\cite{lee2021android} explores the variance-bias trade-off of malware
clustering on 134,698 Android apps. By comparison, our work focuses on the
measurement study of large-scale Windows malware collections. Our goal is
to characterize the applicability and limits of ML-driven malware
classifiers for practical use.  In \cite{motif}, Joyce et al. built a
multi-family dataset containing 3,095 malware samples collected from
454 families. This work offers the most diversified coverage over different
malware families in public malware datasets with manually verified labels.
Interestingly, this dataset has a highly skewed distribution over the
number of malware samples per family. Over half of the families contain
less than 5 samples per family, which poses a few-shot learning challenge
to ML-driven malware classification. 
Our study tried to mimic this distribution to assess the impact of the skewed distribution of malware samples over the accuracy of the trained ML-based classifiers.
We also compare the impact of the skewed distribution and
that of varying malware coverage regarding classification accuracy. 
The empirical study helps identify the limits of ML-based classification methods in practical malware analysis. 



\section{Final Recommendations} 
\label{sec:conclusions}

The goal of this work was to understand the key factors that influence the
performance of ML models for malware detection and family
classification. 
Based on our experimental results,
we can draw some general recommendations on the use of ML for
malware classification:

\begin{enumerate}[leftmargin=0.4cm]
\item[1.] 
 \revision{Ideally, experiments on malware classification (both binary and family) should
    be performed on hundreds of different families, each containing a sufficient
    and balanced number of samples. However, this is often difficult to 
    achieve in the malware field. Thus, we believe the contribution of our paper
    is not to simply re-state this obvious finding, but to provide for the first time
    a quantitative assessment of the impact of the lack of these
    characteristics on the classification results. For instance, we show that
    classifiers trained on a few
    families (like the ones using the popular Microsoft dataset) can provide
    misleadingly high accuracy scores while experiments conducted on unbalanced datasets
    tend to generalize poorly when tested over different distributions.\\
    Our findings can also be used to better understand and compare results reported
    in previous studies. For example, our results show that a family
    classifier with a F1 score of 0.89 over 600 families is likely
    better than a classifier with a score of 0.93 on 30 families.}



\item[2.] Static features dominate detection and classification of samples from
   \textit{known} families, by relying on signature-like information extracted
   from sequences of bytes and opcodes. Packing, in its current widespread implementation,
   does not seem to have a considerable negative effect on this.
   The addition of dynamic features, which are much more time-consuming and
   error-prone to extract, has only a marginal impact on the classification accuracy
   and therefore its use should be carefully considered if the goal is to detect
   known families.
   However, static features are unable to capture samples from
   \textit{unknown} families, where instead models based on dynamic
   behavior show a better ability to generalize.
   \revision{Therefore, our findings suggest that \emph{today} static features alone 
       are sufficient for family classification, but a combination of static and dynamic
       features is probably preferable for binary classification.}

\item[3.] The performance of \textit{all} ML models drop drastically when tested on OOD samples.
    \revision{While the feature completeness and the regular update of the training data to cover new malware families are key to obtaining good classification accuracy,
   both of them are difficult to achieve in the real world. It is due to the data-driven nature of ML-based classification mechanisms. 
   The quality and coverage of training data play a core role in determining the classification performance. 
Beyond improving the quality of training data, our experiments suggest that the inclusion of dynamic features into the classification task can be used to alleviate the impact of the OOD issue. 
More specifically, we show that using dynamic features still allows us to successfully flag suspicious previously-unseen malware samples, even if with less accuracy and higher false positive rates in binary and family classification tasks.
}
\end{enumerate}

\revision{Our work opens several directions for future work. 
For example, we would like to explore 
how to mitigate the impact of missing features in dynamic analysis, 
e.g., through feature selection.
We also plan to analyze the reasons behind hard-to-detect families, 
which could be due to 
custom packers, 
benign functionality in the malware, 
generic families that cover different malware, or 
other reasons.}  


\section*{Acknowledgment}
We would like to thank the anonymous reviewers for their constructive feedback.
We are also grateful to Dan Marino and Chris Gates for their help with the datasets.
This work has benefited from two
government grants managed by the National Research Agency under France
2030 with the references "ANR-22-PECY-0007" (DefMal) and "ANR-22- PECY-0008" (Superviz), 
and it was supported by the European Research
Council (ERC) under the European Unions Horizon 2020 research and 
innovation programme under grant agreement
771844 (BitCrumbs).
This work has also been partially supported by the 
Spanish Ministry of Science and Innovation 
through grants TED2021-132464B-I00 (PRODIGY) and
PID2022-142290OB-I00 (ESPADA).
These projects are co-funded by the European Union
EIE and NextGeneration EU/PRTR Funds.

\bibliographystyle{ACM-Reference-Format}
\bibliography{biblio}
\appendix
\section{Appendix}

\subsection{Impact of Missing Dynamic Feature Values}
\begin{figure}[h]
\centering
\includegraphics[width=\columnwidth]{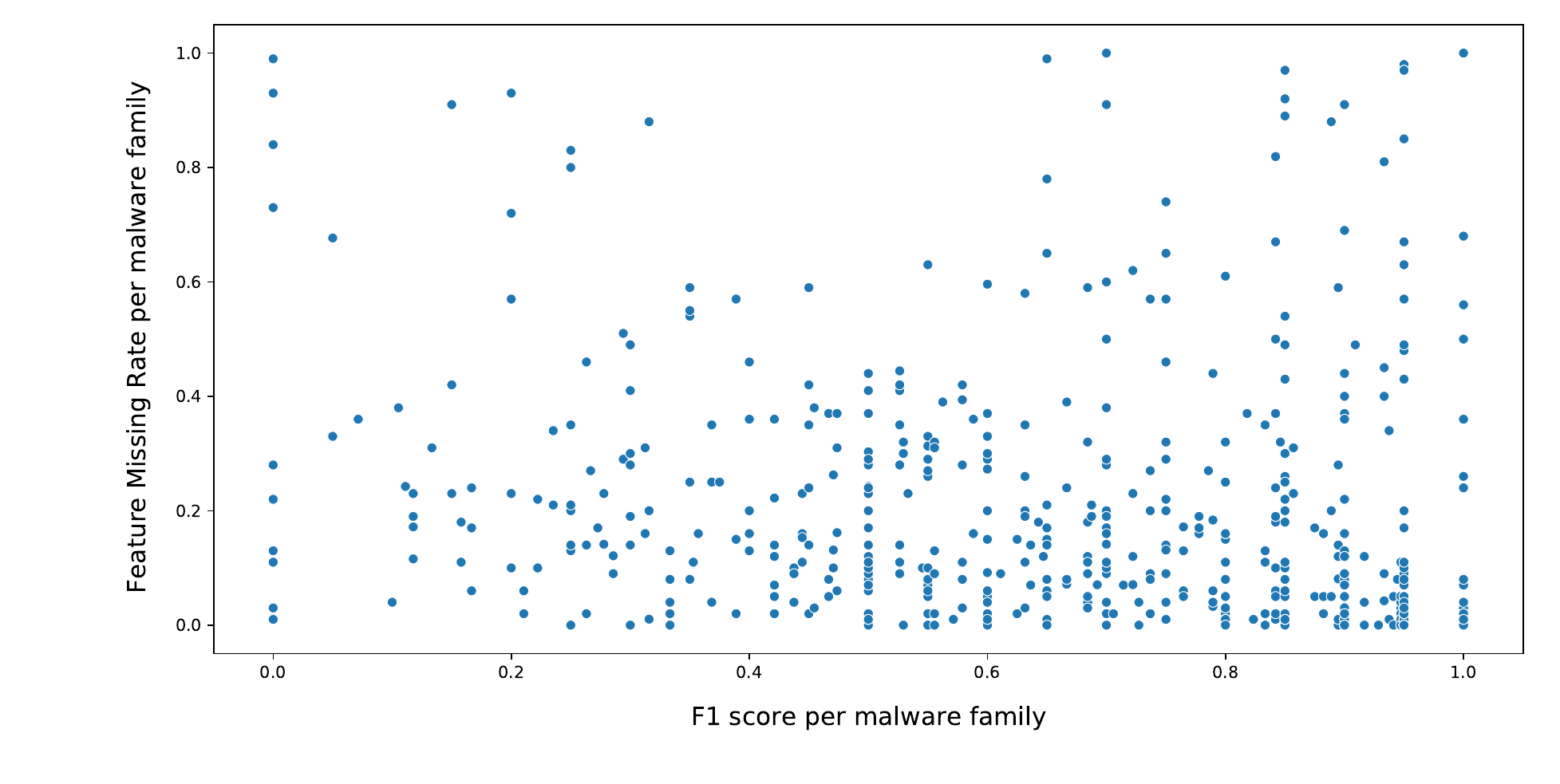}
\caption{F1 score for family classification using dynamic features 
versus Feature Missing Rate (FMR) for the family.}
\label{fig:f1vsfmr}
\end{figure}

Figure~\ref{fig:f1vsfmr} compares the family-wise classification accuracy (F1 score) 
for a family with its FMR for the family classification task. 
The figure shows that a lower FMR tends to produce higher F1 scores and 
vice versa. 

\subsection{IG distribution for \emph{s-bytegrams}, and \emph{s-opcodegrams} classes}\label{sec:IG}
\begin{figure}[h]
  \begin{subfigure}{.23\textwidth}
	\includegraphics[width=1\columnwidth]{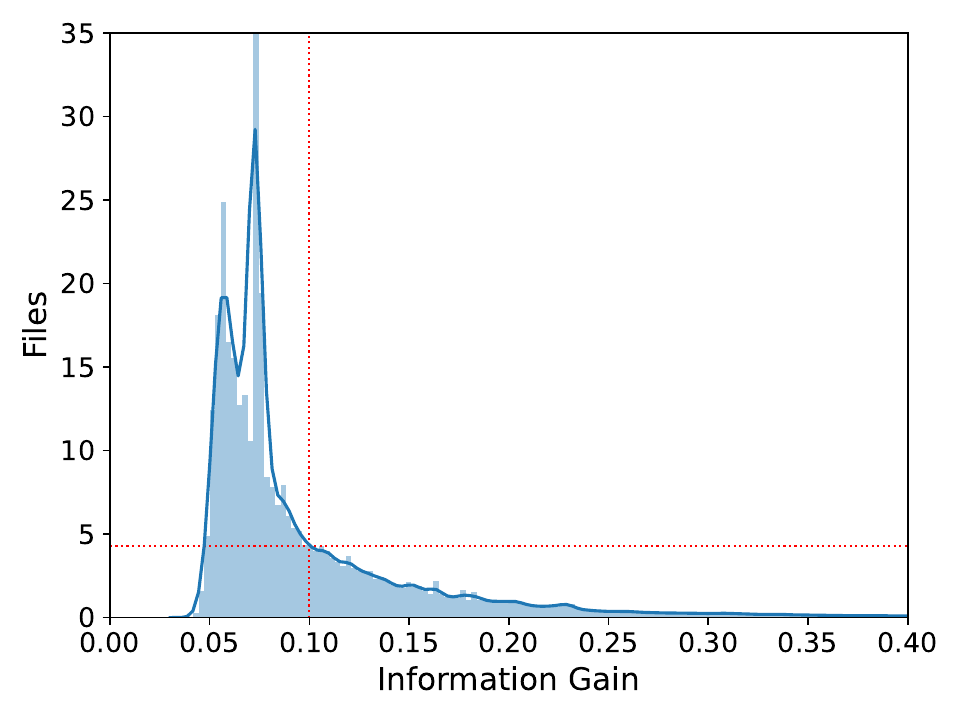}
	  \caption{s-bytegrams}
	\label{fig:IG_ngrams}
  \end{subfigure}
  \hfill
  \begin{subfigure}{.23\textwidth}
	\includegraphics[width=1\columnwidth]{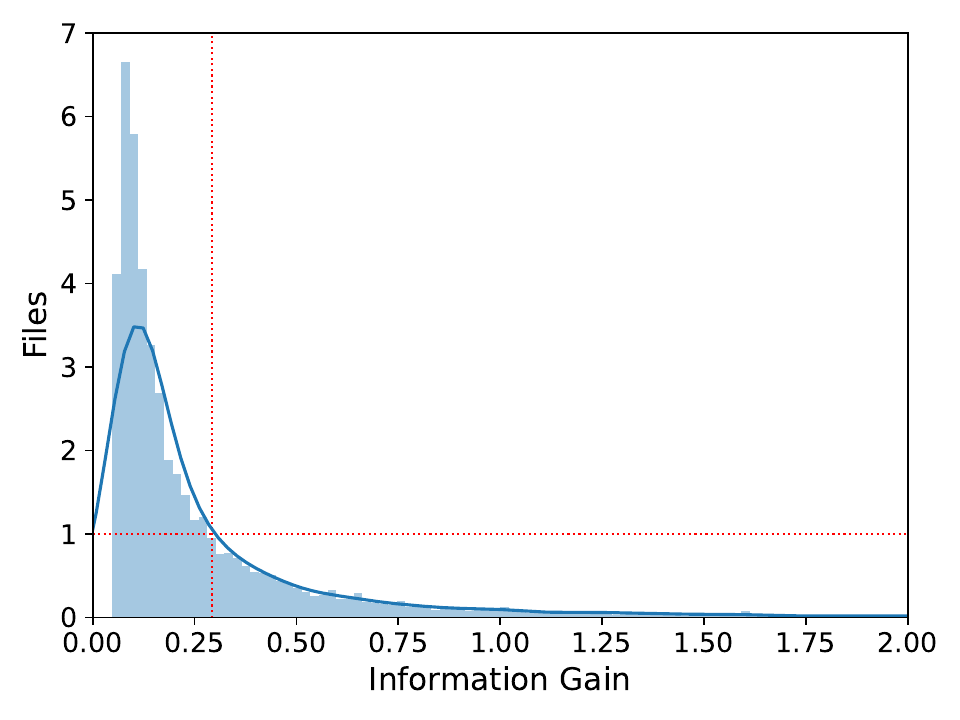}
	  \caption{s-opcodegrams}
	\label{fig:IG_opcodes}
  \end{subfigure}
	\caption{Distribution of number of \emph{s-bytegrams} and \emph{s-opcodegrams} based on their Information Gain value.}
	\label{fig:IG}
\end{figure}

\subsection{Feature list for \emph{s-headers} and \emph{s-sections}}
\label{sec:peHeaders-peSections}
\begin{table*}[!h]
\small
\centering
\caption{Summary of the features extracted from PE headers}
\label{tbl:staticFeatures_peHeaders}
\begin{tabular}{lll}
\hline
\textbf{Feature name} & \textbf{Header} & \textbf{Description} \\
\hline
	ImageBase 				&	Optional	& The address of the memory mapped
	location of the file\\
	AddressOfEntryPoint		&	Optional	& The address where the loader will
	begin execution\\
	SizeOfImage 			&	Optional	& The size (in bytes) of the image
	in memory\\
	SizeOfCode 				&	Optional	& The size of the code section\\
	BaseOfCode 				&	Optional	& The address of the first byte of
	the entry point section\\
	SizeOfInitializedData 	&	Optional	& The size of the initialized data
	section/s\\
	SizeOfUninitializedData	&	Optional	& The size of the uninitialized data
	section/s\\
	BaseOfData				&	Optional	& The address of the first byte of
	the data section\\
	SizeOfHeaders 			&	Optional	& The combined size of the MS-DOS
	stub, PE headers, and section headers\\
	SectionAlignment 		&	Optional	& The alignment of sections loaded
	in memory\\
	FileAlignment 			&	Optional	& The alignment of the raw data of
	sections\\
	NumberOfSections 		&	COFF		& The number of sections\\
	SizeOfOptionalHeader 	&	COFF 		& The size of the optional header\\
\hline
	Characteristics bit 	&	COFF		& 16 Boolean values - one for each bit of
	the Characteristics bit~\cite{microsoftPE} \\

\hline
\end{tabular}
\end{table*}

\begin{table*}[!h]
\small
\centering
	\caption{Summary of the features extracted from PE sections.
	\emph{Processed resources} and \emph{Processed sections} reflect max, mean, and min values computed among all the resources and sections in the PE file. 
	The last block of features are computed for each $i^{th}$ section and for the section containing the binary entry point, and then only features that show variability and are present in more than 1\% of the
	samples retained.}
\label{tbl:staticFeatures_peSections}
\begin{tabular}{ll}
\hline
\textbf{Feature name} 					\textbf{Description} \\
\hline

	Processed\_resources\_nb 					&	Resource number in the PE		\\
	Processed\_resourcesMaxEntropy  			&	Max Shannon entropy among resources	\\
	Processed\_resourcesMaxSize 				&	Max size among resources		\\
	Processed\_resourcesMeanEntropy 			&	Mean Shannon entropy among resources	\\
	Processed\_resourcesMeanSize    			&	Mean size among resources		\\
	Processed\_resourcesMinEntropy  			&	Min Shannon entropy among resources	\\
	Processed\_resourcesMinSize 				&	Min size among resources		\\
\hline
	Processed\_sectionsMaxEntropy   			&	Max Shannon entropy among sections		\\
	Processed\_sectionsMaxSize  				&	Max size among sections		\\
	Processed\_sectionsMaxVirtualSize   		&	Max virtual size among sections	\\
	Processed\_sectionsMeanEntropy  			&	Mean Shannon entropy of all the sections		\\
	Processed\_sectionsMeanSize 				&	Mean size of all the sections	\\
	Processed\_sectionsMeanVirtualSize  		&	Mean virtual size of all the sections		\\
	Processed\_sectionsMinEntropy   			&	Min Shannon entropy of all the sections		\\
	Processed\_sectionsMinSize  				&	Min size of all the sections	\\
	Processed\_sectionsMinVirtualSize   		&	Min virtual size of all the sections		\\
\hline
	Section\_i\_exists						&	True if the $i^{th}$ section exists in the binary									\\
	Section\_i\_name\_is\_standard			&	True if the $i^{th}$ section has a standard name~\cite{microsoftPE}	\\
	Section\_i\_size						&	Size of the $i^{th}$ section \\
	Section\_i\_phisicalAddress					&	The starting address of the $i^{th}$ section in the file	\\
	Section\_i\_virtualSize					&	The total size of the $i^{th}$ section in memory \\
	Section\_i\_entropy						&	The Shannon entropy of the $i^{th}$ section \\
	Section\_i\_numberOfRelocations			&	The number of relocation entries in the $i^{th}$ section \\
	Section\_i\_pointerToRelocations		& 	The address of the first byte of the relocation entries in the $i^{th}$ section \\
	Section\_i\_characteristics bit 		& 	32 Boolean values - one for each bit of the section Characteristics bit~\cite{microsoftPE} \\

\hline
\end{tabular}
\end{table*}

\end{document}